\documentclass[final,5p,times,twocolumn]{elsarticle}

\usepackage{amssymb}

\usepackage[default]{lato}
\usepackage[utf8]{inputenc}
\usepackage{graphicx}
\usepackage{multirow}
\usepackage{multicol}
\usepackage{subfigure}
\usepackage{hyperref}

\usepackage{xcolor}
\usepackage{listings}
\usepackage{color}
\usepackage{afterpage}

\definecolor{dkgreen}{rgb}{0,0.6,0}
\definecolor{gray}{rgb}{0.5,0.5,0.5}
\definecolor{mauve}{rgb}{0.58,0,0.82}

\lstdefinestyle{quote}{
  aboveskip=1mm,
  belowskip=1mm,
  basicstyle={\normalsize\color{blue}},
  columns=fullflexible,
  breaklines=true,
  breakatwhitespace=true,
  showstringspaces=false,
  breakindent=0pt,
  captionpos=b
}

\usepackage[most]{tcolorbox}

\newtcolorbox{mybox}{
    colback=blue!5!white, 
    colframe=blue!75!black, 
    coltext=black, 
    boxsep=0pt, 
    arc=4mm, 
    breakable,
    fontupper=\normalsize, 
    fonttitle=\normalsize 
}

\usepackage{float}
\floatstyle{boxed} 
\newfloat{List}{t}{lop}
\usepackage{rotating}

\definecolor{quotationcolour}{HTML}{F0F0F0}
\definecolor{quotationmarkcolour}{HTML}{1F3F81}

\usepackage[noabbrev,capitalise]{cleveref}

\journal{Journal of Systems and Software}

\begin{document}

\begin{frontmatter}



\title{An Empirical Study on Governance in Bitcoin's Consensus Evolution}


\author[NTNU1]{Jakob Svennevik Notland}
\author[NTNU2]{Mariusz Nowostawski}
\author[NTNU1]{Jingyue Li}

\affiliation[NTNU1]{organization={Department of Computer Science, Norwegian University of Science and Technology},
            addressline={Sem Sælandsvei 9}, 
            city={Trondheim},
            postcode={7491}, 
            state={Trøndelag},
            country={Norway}
}

\affiliation[NTNU2]{organization={Department of Computer Science, Norwegian University of Science and Technology},
            addressline={Teknologiveien 22}, 
            city={Gjøvik},
            postcode={2815}, 
            state={Innlandet},
            country={Norway}
}

\begin{abstract}
Blockchain systems run consensus rules as code to agree on the state of the distributed ledger and secure the network. Changing these rules can be risky and challenging. In addition, change can be contentious, and getting all necessary participants to agree on a change could be tedious.
Arguably, Bitcoin has seen centralisation tendencies in mining and development. However, how these tendencies influence governance processes of consensus evolution has received minimal community and academic attention.
We explore how blockchain systems' evolution and governance of consensus rules are intertwined from socio-technical aspects. Our study analyses the governmental structures in blockchain by looking into Bitcoin's history. We investigate processes of changing consensus rules through a grounded theory analysis comprising quantitative and qualitative data from 34 consensus forks in Bitcoin and Bitcoin Cash.
This study explores how decentralisation and system evolution governance principles unfold in practice. 
In contrast to existing studies, we revealed that the centralisation tendencies among miners and developers have no noticeable impact on the decentralisation of decision-making power.
Furthermore, the centralisation tendencies do not affect decision-making for consensus evolution governance in the same way as they facilitate consensus attacks, such as 51\% attacks or selfish mining. 
We also discovered that consensus evolution governance is constrained by the technicalities of a change and the deployment techniques. Consequently, the decision-making power to make permanent consensus rule changes lies with the miners, constructing the longest chain, restricted by the different deployment techniques and dependence on user adoption.  


\end{abstract}


\begin{highlights}
\item Show how blockchain evolution governance contrasts with classic software evolution.
\item Depict roles in blockchain evolution governance and their decision-making power.
\item Show how Nakamoto's envisaged decentralised power structure works in practice.
\item Demonstrate how fork deployment and evolution governance are intertwined.
\item Show how techniques of evolution evolve to fit the needs of consensus change.
\end{highlights}


\end{frontmatter}


\section{Introduction}

Consensus rule changes in public permissionless blockchains such as Bitcoin are 
challenging~\cite{BIP-0050,invalidBlocks,inflationVulnerability}, and the most trivial change could cause suspended services~\cite{suspendedService},
lost mining revenue~\cite{invalidBlocks}, double-spending~\cite{OKPay}, and replay attacks~\cite{kiffer2017stick}.
However, reaching a consensus to adopt a change could prove more challenging than implementing the change itself~\cite{P2SHWar,roadToSegwit}.  
The process to extend the blockchain (e.g., Bitcoin) was originally described as a majority decision, i.e.,  ``essentially one-cpu-one-vote"~\cite{nakamoto2019bitcoin}. Although we have long since moved on from CPU voting to dedicated hardware, the theory still implies that the majority of computational power determines the evolution of the blockchain and consensus. 

Related work suggests that repository maintainers in Bitcoin act as centralising factors because only a small group of highly skilled developers can change the rules of the system~\cite{de2016invisible, parkin2019senatorial}.
On the other hand, we hypothesize that other factors affect evolution governance and that these aspects should be explored further.
In addition, pool centralisation can be a worry regarding majority attacks (i.e., 51\% attacks)~\cite{beikverdi2015trend, romiti2019deep}.
Similarly, majority attacks could affect the consensus evolution in blockchain, although that has not been considered in related work.
These concerns raise questions on how consensus evolves in practice and how it is governed. Thus, we aim at understanding  ``\textit{How are the attributes of the Bitcoin network utilised as governmental instruments in consensus evolution?}"

In contrast to the traditional evolution of open-source software~\cite{guzzi2013communication, sharma2021extracting}, Bitcoin is an immutable and consistent distributed ledger relying on every stakeholder to participate in the evolution of the system itself.
Even if a small group of developers decide to adopt consensus changes on a repository level, they will need support from the network to deploy changes.
In theory, the decision-making power to change consensus rules relies on the adoption of miners and the majority of computing power. Furthermore, a proposal's success depends on network-wide consensus and user adoption.
To better understand consensus evolution governance in blockchain, we focus on answering the following research questions:

\begin{itemize}
    \item \textbf{RQ 1.} Which roles and actors can impact changes to consensus?
    \item \textbf{RQ 2.} What is the decision-making process for consensus changes?
    \item \textbf{RQ 3.} How is decision-making power distributed among the different roles in consensus evolution?
 \end{itemize}

To answer the research questions, we conduct an empirical study to explore consensus evolution governance in blockchain using Bitcoin as a case study.
We chose Bitcoin because it has a long, well-documented operational history and many consensus change events. We gathered data from Bitcoin's decision-making process, showing how off-chain and on-chain coordination drive evolution.
Furthermore, we analysed qualitative data from developer communication channels and quantitative data from the Bitcoin ledger using Strauss' Grounded Theory (GT)~\cite{corbin2014basics} approach.
Our analysis encompasses 34 consensus forks in Bitcoin Core (BTC) and Bitcoin Cash (BCH) from the launch in 2009 until 2022. We mainly collect samples from developer channels, e.g., emails, forums, GitHub, and Internet relay chat (IRC). Our results are summarised as follows:

\begin{itemize}
  \item \textbf{RQ 1.} We show how roles in blockchain governance closely resemble the tripartite system's separation of power~\cite{tripartite} where developers act as~\textit{legislature}, miners are~\textit{executive}, and full nodes are~\textit{judiciary} entities of power. The difference in a permissionless blockchain such as Bitcoin is that actors can choose which system they want to be part of, and any actor may simultaneously resemble multiple or all the roles. 
  \item \textbf{RQ 2.} We identify the steps in the governance process and how they intertwine with consensus change deployment. The findings reveal how governance processes of permissionless blockchains, such as Bitcoin, differentiate themselves from classical software evolution and show that consensus change implementation and deployment are dependent and the community must collaborate and coordinate in a decentralised manner to adopt and activate the changes. 
  
  \item \textbf{RQ 3.} We found that, despite potential developer and pool centralisation, the decision-making power of the economic majority, miners, full nodes, and users are decentralising factors in the governance process. Furthermore, we observed a limitation of Nakomoto's decentralised power structure: only a fraction of the participants (the developers) can judge the soundness of a change. The remaining roles (miners, full nodes, and users) merely speculate on the technical and economic implications of the code changes.
\end{itemize}

The findings lead to the following contributions:
\begin{enumerate}
    \item We provide novel formalizations of the decision-making processes and capabilities for actors and roles involved in evolution governance.
    \item We bring novel insights that blockchain evolution governance is constrained by the technicalities of a change, i.e.,~compatibility with older implementations,  and the deployment technique, i.e., conditions for activation, which differentiate the evolution governance of permissionless blockchain from classic software.
    \item Compared to the state-of-the-art studies on blockchain and open-source software evolution, which focus solely on evolution governance without considering how deployment techniques change over time, we revealed that deployment techniques in blockchain are evolved to fit the needs of consensus evolution and explained how they are evolved. 
\end{enumerate}





The remainder of this paper is structured as follows: 
\Cref{backrgound} introduces the background on blockchain consensus and its evolution. \Cref{method} describes the research design and implementation. \Cref{results} presents the results, describing consensus evolution governance. \Cref{discussion} discusses the results before the conclusion in \Cref{conclusion}.

\section{Background} \label{backrgound}




\subsection{The Bitcoin consensus protocol}

\begin{figure}
  \centering
  \includegraphics[width=1\linewidth]{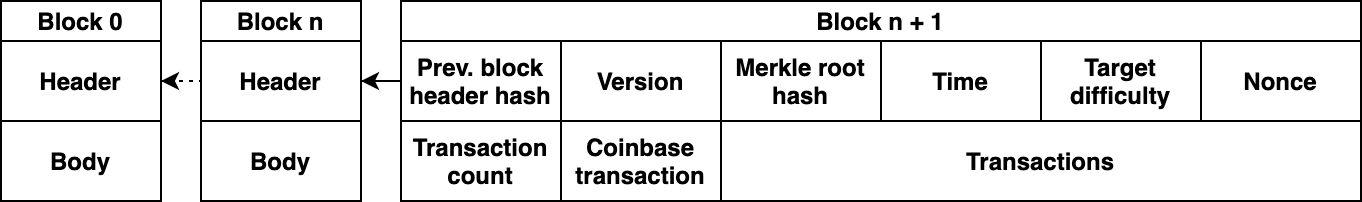}
  \caption{Bitcoin blocks in a blockchain showing the content of the header and the body. A new block must comply with the consensus rules, e.g., only include valid transactions and find a nonce, creating a blockhash that meets the target difficulty. The figure was first presented in~\cite{self}.}
  \label{fig:block}
\end{figure}

Bitcoin is ``a peer-to-peer electronic cash system"~\cite{nakamoto2019bitcoin} consisting of a chain of blocks containing transaction history, as illustrated by~\cref{fig:block}.
Any new block must abide by the consensus rules to be accepted by the network.
Miners attempt different values for the nonce variable in a brute-force manner to produce an SHA-256 hash based on the entire block.
The miners find a valid nonce for the block header when the resulting hash meets the target difficulty. The difficulty indicates that the resulting hash must have a certain number of leading zeros. 
This mechanism is known as Proof-of-Work (PoW), first proposed to prevent email spam~\cite{dwork1992pricing} and later applied for cryptocurrencies~\cite{jakobsson1999proofs}.
In the case of Bitcoin, PoW prevents Sybil attacks~\cite{douceur2002sybil} and ensures immutability. These principles are preserved by the difficulty adjustment algorithm~\cite{nakamoto2019bitcoin}, ensuring that the network will produce blocks with an average rate of around ten minutes. Miners must comply with further consensus rules for block and transaction validity. Miners are incentivised to follow these rules by collecting fees and the block reward generated in the {\tt coinbase transaction}.

\begin{figure}
  \centering
  \includegraphics[width=1\linewidth]{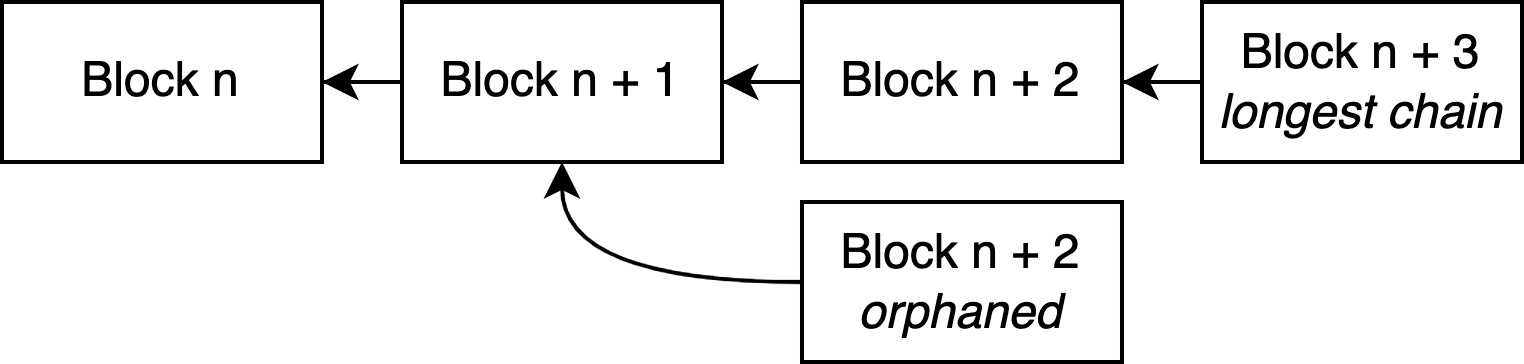}
  \caption{Choosing the longest chain during a collision. Miners choose to extend one of the competing blocks at height (n + 2). As soon as another block (n + 3) is found, that chain becomes the longest chain. The other block becomes orphaned. The figure was first presented in ~\cite{self}.}
  \label{fig:collision}
\end{figure}

Block collisions occur when two valid blocks are produced at the same height at approximately the same time. These blocks can contain different or even conflicting transactions.
Collisions are resolved by {\it the longest (valid) chain rule} as specified in Nakamoto's whitepaper: ``The majority decision is represented by the longest chain, which has the greatest proof-of-work effort invested in it" \cite{nakamoto2019bitcoin}. The notion of a {\it valid} chain is vital because validness is subjective from the implementation's point of view.
For instance, BCH miners do not perceive the BTC chain as valid, albeit the longest chain. 
The longest chain rule ensures that block collisions are quickly resolved during regular operation. Whenever another block is appended on top of one of the colliding blocks, it becomes the longest chain. This behaviour implies the slight chance of {\it orphaning} (discarding) blocks, along with the corresponding transactions as shown in~\cref{fig:collision}. 
To prevent financial loss from orphaned blocks, one should wait until the block containing the relevant transaction has at least six blocks built on top, i.e., six confirmations~\cite{rosenfeld2014analysis}. 
Collisions happen naturally, by an attack~\cite{sayeed2019assessing}, or by inconsistent consensus validation~\cite{invalidBlocks}. 

\subsection{Bitcoin consensus evolution}

\begin{figure}[t]
    \centering
    \includegraphics[width=1\linewidth]{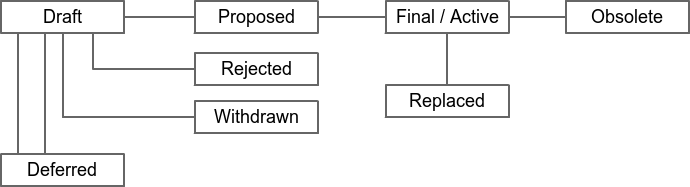}
    \caption{BIP-2: Figure from "BIP process, revised"~\cite{bips}. The BIP process describes how contributors can formalise proposals and the proposal lifecycle before potential deployment and activation in the Bitcoin network.}
    \label{fig:bip}
\end{figure}

The process for Bitcoin Improvement Proposals (BIPs), seen in
~\cref{fig:bip}~\cite{bips}, was adopted from Python Enhancement Proposals (PEPs)~\cite{peps}, which were based on Request for Comment (RFC)~\cite{rfc}.
These processes formalise how to handle proposals before eventual deployment. In the early days of Internet protocols, governance relied on "rough consensus and running code" to settle disputes~\cite{russell2006rough}.
In contrast, Bitcoin is an immutable and consistent distributed ledger shifting from rough to uniform consensus to secure its integrity and value. Even if a group of developers decide to adopt consensus changes on a repository level, they need network support for deployment.

An inherent ideology within Bitcoin and especially BTC (in comparison to BCH) indicates what kind of changes are viable or cause contention. The quote below from the creator(s) can be seen as a cornerstone of this ideology.

\begin{lstlisting}[style=quote, caption={Satoshi 2010-06-17 Forum, ID: 195}, label=cornerStone]
The nature of Bitcoin is such that once version 0.1 was released, the core design was set in stone for the rest of its lifetime.
\end{lstlisting}

The statement from Nakamoto explains that the implementation and the original specification~\cite{nakamoto2019bitcoin} define Bitcoin's fundamental behaviour.
It also highlights the importance of non-disruptive changes, no matter how insignificant they may seem.
Additionally, changes that alter consensus rules in Bitcoin should allow backward compatibility such that legacy nodes can accept new behaviour based on the original specification, keeping the network consistent.
Our work distinguishes between Bitcoin, i.e., the idea proposed by Nakamoto, and BTC and BCH, which are different implementations of Bitcoin with varying approaches to consensus evolution governance. BCH was created as a fork in reaction to deploying the controversial segregated-witness fork in BTC~\cite{roadToSegwit}. The governance processes related to segregated-witness are particularly interesting for our study because of the community contention and multiple consensus change events associated with a single feature.

When referring to \textit{rules} and \textit{rule changes} in this paper, we are referring to changes in blockchain consensus rules. 
A blockchain node or miner may validate a subset of those rules, while \textit{full nodes} validate based on all the rules. 
Sometimes, backwards-incompatible changes are needed if the implementation does not work as intended. Such a change could be required to prevent exploits or allow for the adaption and survival of the system. As shown in~\cref{uptight}, the Bitcoin community apply changes cautiously and, especially, consensus changes because bugged or ill-intended code may disrupt the system's integrity and stability. 


\begin{lstlisting}[style=quote, caption={2011-08-10 Gavin Andresen Email: bitcoin-dev}, label=uptight]
5. Testing. I don't have time to personally test every PULL request, but if a pull involves more than trivial code changes I'm not going to pull it unless it has been thoroughly tested.  We had a very good rule at a company I used to work for-- programmers were NOT allowed to be the only ones to test their own code. Help finding money and/or people for a dedicated "core bitcoin quality assurance team" is welcome. More unit tests and automated testing is also certainly welcome.

If this was open source blogging software I'd be much less uptight about testing and code review and bugs. But it's not, it is software for handling money.
\end{lstlisting}

The BTC community has a conservative approach to consensus changes. \Cref{fig:timeline} shows that consensus changes are rarely deployed and can often cause disruption in the blockchain or the community.

\subsection{Bitcoin consensus change deployment}

Consensus changes in BTC have mainly been deployed as backwards-compatible soft forks~\cite{princetonBlockchain}. 
The fork type relies on the compatibility between old and new versions. The fork types combine with three deployment strategies, user-activation, miner-activation and emergency-activation, to form deployment techniques~\cite{self}. BTC has mainly been using deployment techniques, such as IsSuperMajority (ISM)~\cite{BIP-0065}, and later, the more mature miner-activated soft fork (MASF)~\cite{BIP-0009}.
These techniques require miners to signal on-chain readiness before executing a coordinated activation depending on a threshold.
Other well-known techniques are user-activated soft fork (UASF)~\cite{BIP-0148} or BCH's preferred user-activated hard fork (UAHF)~\cite{UAHF}.
These techniques activate based on a chosen block height (flag day), and changes apply regardless of the adoption percentage. All the deployment techniques are explained in previous work~\cite{self}, where we argue that the user-activated forks value new functionality over consistency in the network, carrying higher consensus failure risk~\cite{self}.

As opposed to using popular terms such as {\tt soft fork} and {\tt hard fork} to describe different kinds of changes, we use Zamyatin et al.'s. notion of \textit{reduction}, \textit{expansion}, and \textit{bilateral} forks~\cite{zamyatin2018wild} for a more accurate description.

A {\bf reduction fork} (AKA {\tt soft fork}) reduces the set of valid actions in a blockchain, and legacy implementations will accept all actions under the new validity set. Reduction forks are backwards-compatible.

An {\bf expansion fork} (AKA {\tt hard fork}) expands the set of valid actions in a blockchain, and the new implementation will accept all actions in the legacy validity set and more. Expansion forks are not backwards-compatible.

A {\bf bilateral fork} (AKA {\tt hard fork}) creates a conflicting validity set where all previously valid actions are invalid according to the fork, and the fork is entirely invalid according to the legacy rules. Bilateral forks are both backwards- and forwards-incompatible

\section{Research design} \label{method}

Related work raises concerns regarding Bitcoin centralisation in development~\cite{parkin2019senatorial,de2016invisible} and mining~\cite{romiti2019deep, beikverdi2015trend}.
In contrast, we hypothesise that these aspects have a low impact on the power distribution in the governance of Bitcoin and permissionless blockchains in particular. We assume this because consensus mechanisms and social constructs restrict blockchain consensus change governance and deployment capabilities~\cite{self}.


From these aspects, we explore blockchain's critical yet less understood aspect: the interplay between decentralisation and governance.
Blockchain's innovation lies in its decentralised structure, offering autonomy and system resilience absent in traditional, central-authority-based systems.
However, the original design of Bitcoin, predicated on majority decision-making, does not necessarily mirror the practical nuances of governance encountered in its real-world application. This discrepancy underscores the need for empirical investigations. 
\subsection{Research method}

This work builds on and extends our previous work investigating runtime evolution of consensus rules and characterises deployment techniques~\cite{self}, which identifies nine features used for deployment and how they resemble nine different techniques.
Therefore, the qualitative samples considered are the same, while this work adds new analysis and focuses on the governmental aspects of consensus evolution.
Furthermore, we introduce new data from on-chain governance traces from block data because they provide new insights into the governance processes.

We chose Strauss' grounded theory (GT) approach, which is suited for studies with predefined research questions~\cite{stol2016grounded}.
The GT approach is iterative and recursive, and the researchers must go back and forth until reaching theoretical saturation, i.e.,~when new samples stop expanding the developing theories.
The observations in our study are covert~\cite{oates2006researching}, where we get the most authentic experience of how the actors conduct a process. 
Covert observations can be ethically questionable. However,  Bitcoin's public archives were created to provide transparency and accountability.
Therefore, the actors should expect their actions to be considered in research. We represent the cited actors by their chosen aliases without any effort to reveal the person behind or interconnect different aliases.



\begin{figure*}[t]
    \centering
    \includegraphics[width=\textwidth]{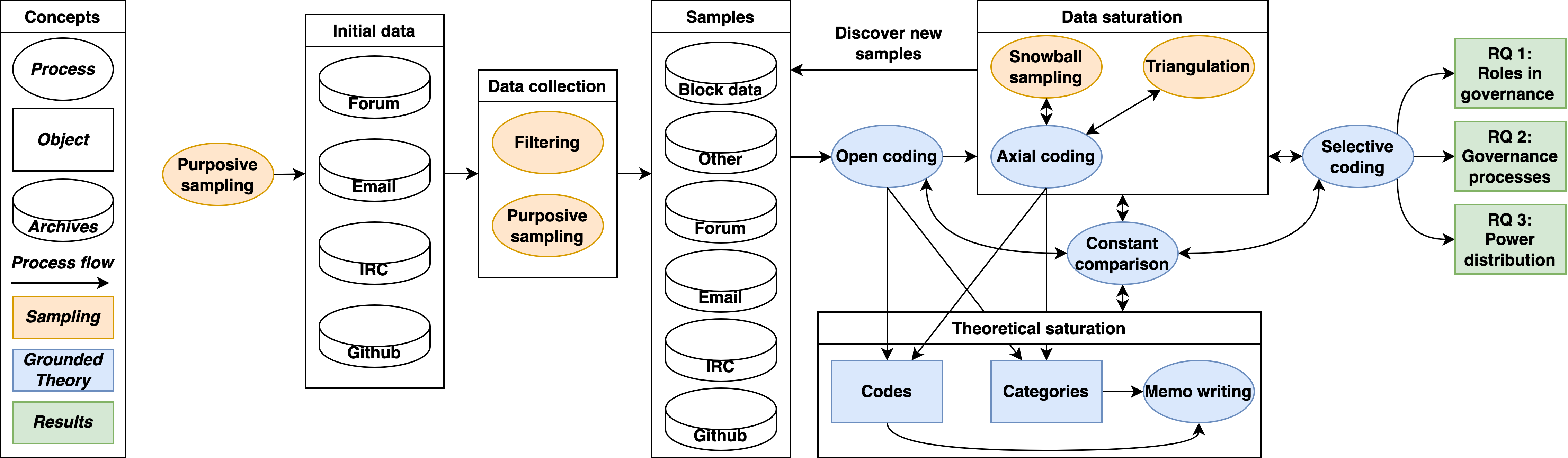}
    \caption{In the research method implementation, oval shapes represent processes, lines indicate the flow of these processes, rectangles denote data objects, and cylinders signify archives. Different colours highlight whether the concepts relate to data sampling or grounded theory analysis.}
    \label{fig:method}
\end{figure*}

This study started with sampling~\cite{oates2006researching} from Bitcoin Core's development archives. The samples were collected by purposive sampling (\cref{sec:purposive}) and filtering (\cref{sec:filtering}) those relevant to consensus changes. These samples were sorted using flexible coding~\cite{deterding2021flexible} (\cref{codesCategories}).
Further, we applied snowball sampling and triangulation \cite{corbin2014basics} (\cref{snowballing}) to avoid limitations from the initial samples. A graph of the research method is outlined in \cref{fig:method}.

\subsubsection{Purposive sampling} \label{sec:purposive}

Bitcoin Core's consensus changes are listed in Bitcoin's Wiki \cite{consensusChanges} and in \ref{dataCollection} \cref{tab:changes}. 
The initially selected archives were the Bitcoin improvement proposals (BIPs)~\cite{bips}, the bitcoin-dev emails \cite{bitcoinEmail}, the Development \& Technical Discussion topic in the Bitcoin forum~\cite{bitcoinTalk}, the code repository (pull requests~\cite{pulls} and issues~\cite{issues}), and from IRC (\#bitcoin-dev and \#bitcoin-core-dev \cite{bitcoinIRC}).
The total sample count is 1700 (available online~\cite{exportedSamples}). One sample corresponds to one proposal, one thread (email, forum and GitHub), or one day of IRC messages. \cref{fig:samplesEvents} shows the distribution of samples based on events, \Cref{fig:samplesSources} depicting the distribution of samples per archive. 

The sample boundary was fluid, so samples from other domains were considered whenever appearing. These domains could be other Bitcoin channels, announcements, blogs, magazines, videos, or other blockchains. We also found traces of consensus governance in the block data, providing further insights into Bitcoin's governance processes.
Data from the blocks show traces of on-chain signals to support different consensus forks. We install the BTC full node implementation \cite{btc} to retrieve the blocks and use {\tt Matplotlib}~\cite{matplotlib} to make graphs representing signals related to consensus forks in BTC. Our script traverses the BTC blockchain from the genesis block until block 737000, including all consensus forks before 2023. 


\subsubsection{Data filtering} \label{sec:filtering}

After surveying consensus change events and BIPs, we applied filtering. Initially, we considered the mailing list. With a compact overview of all the threads, it was viable to traverse through the titles to select relevant samples. 
Due to the large volume and difficulty of navigating, we filtered the forum threads in two stages:
First, threads surrounding the dates of each consensus incident were screened and purposively sampled.
Second, the forum was filtered using the search strings listed in \ref{dataCollection} \cref{tab:changes} with the help of a search tool targeting the Bitcoin forum \cite{ninjastic}.
This filtering approach was also applied to the GitHub search. 


Regarding the IRC channels, we searched for samples by looking for scheduled developer meetings. Only four meetings were found before weekly meetings began in the fall of 2015. Therefore, a search was conducted using a few search strings to identify conversations about consensus changes. These strings were "fork," "chain split," "stuck," and "reorg". Using {\tt grep}~\cite{grep}, the files were traversed all together, showing related sentences. The whole log from each sample was collected whenever indicating relevance.


\subsubsection{Data analysis}
\label{codesCategories}

\begin{figure*}[t]
    \includegraphics[width=\textwidth]{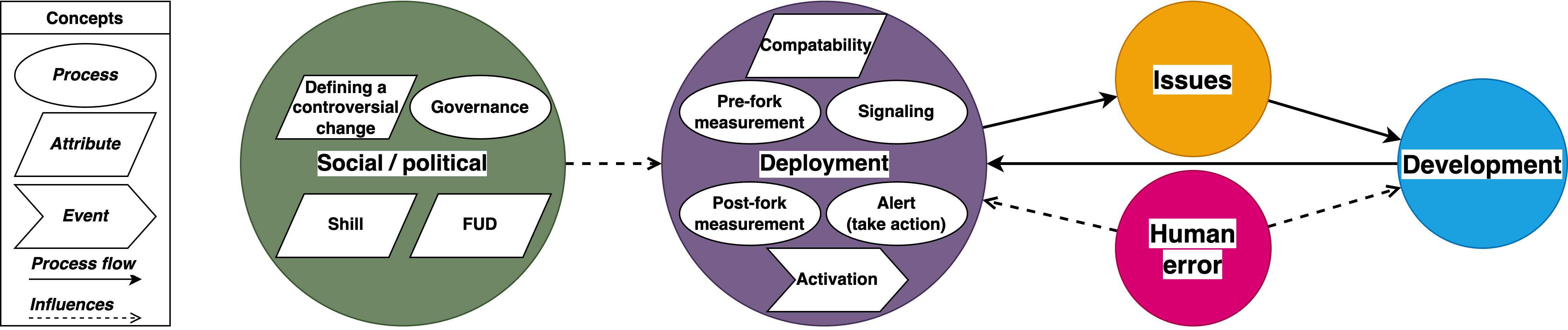}
    \caption{Analytical codes and categories. The relation between the categories indicates a pattern in the cycle of consensus evolution. In the interest of space, we include codes relevant to consensus evolution governance, excluding codes for technical issues, software development and human errors.}
    \label{fig:codesCategories}
\end{figure*}

We utilised flexible coding~\cite{deterding2021flexible}, which is a technique to handle large datasets collected and coded through qualitative data analysis software (QDAS) such as {\tt Atlas.ti}~\cite{atlas}.
To get an overview of the initial data, we applied the indexing approach as a specific form of open coding~\cite{deterding2021flexible}.
In practice, the data was initially indexed on a sample basis to highlight the essence of each sample.
The process of indexing was continuously evaluated as new codes and categories emerged.
Memos were created to understand the correlation between codes, categories, and events. One specific code was used to highlight notable quotes that considerably impacted the results~\cite{deterding2021flexible}. This code always overlaps with other codes and is labelled "aha" to signal the aha experience that these quotes represent. Quotes labelled with this code were continuously revisited and were candidates to present in the article to support our results.


The axial coding phase was conducted by revising, combining, and splitting codes and categories. Some codes were combined with decreasing levels of granularity, and others were split to increase granularity and enhance insights.
This stage also developed patterns and relations within the codes and categories. as shown in~\cref{fig:codesCategories}. We use the codes FUD (Fear, Uncertainty, and Death) and shill to indicate when participants argue against or for a consensus change without providing rational arguments.



At the stage of selective coding, we unfolded and saturated the category of governance to address the research questions. We further elaborated on processes in Bitcoin and other blockchains such as Ethereum and Dash to reflect on the governance structures and their relation to Bitcoin's consensus mechanisms and social constructs.



\subsubsection{Snowball sampling and data triangulation}
\label{snowballing}

The codes, categories, and emerging theories were constantly compared to see whether they made sense regarding the research questions and the objective domain. The initial samples could have led to an adequate theory in this study. However, there was always a possibility that the data set was too narrow. Therefore, we conducted data saturation~\cite{penrod2003discussion,corbin2014basics} to discover additional samples through snowball sampling~\cite{penrod2003discussion} and data triangulation~\cite{corbin2014basics}, providing a rich data set with multiple perspectives. Snowball sampling helped find additional samples and links not part of the initial dataset. For instance, the IRC chat had inconsistencies on the 2015 event (version 0.10.0), where samples were missing from 2015-06-03, 2015-06-04, and 2015-06-05. Other samples revealed how this consensus failure was caused by lazy validation and spy-mining~\cite{bitmex2017spy}.



\section{Results} 
\label{results}

The codes, constant comparison and theoretical saturation based on the codes reveal the system, roles, and processes to develop in a symbiosis of chaos tamed by the underlying blockchain infrastructure. It can be challenging to identify the causalities within the decentralised environment. However, as the analysis commences with codes giving meaning to the data, data triangulation shows the relationships within.
Eventually, this forms a chain of events, shown in the timeline of~\cref{fig:timeline}, showing how the processes of governance evolution are restricted by the underlying blockchain and refined through trial and error.

\subsection{RQ 1: Roles in governance}

\begin{figure}[t]
  \centering
  \includegraphics[width=1\linewidth]{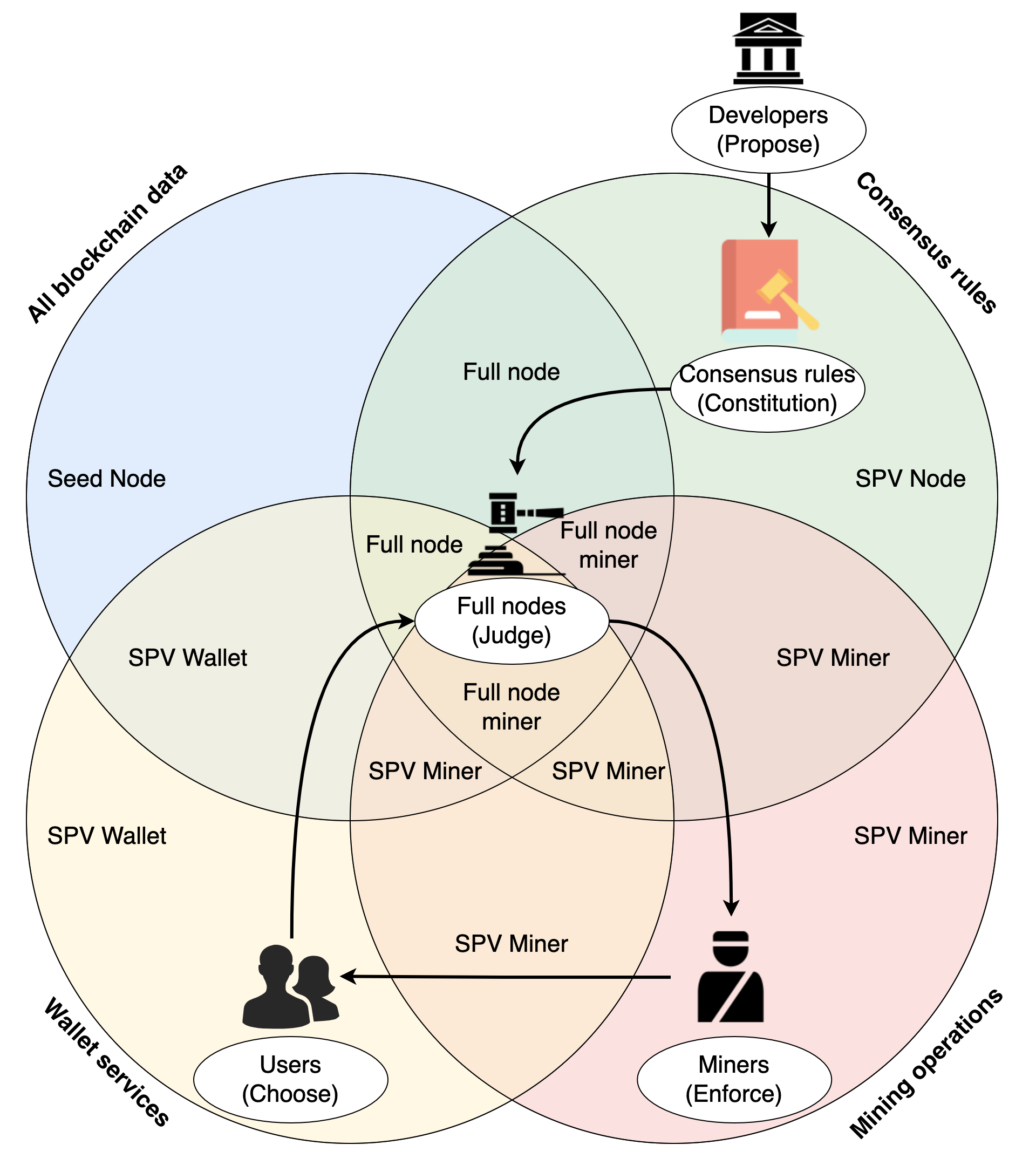}
  \caption{Blockchain node Wenn diagram. The diagram illustrates the takeaway on roles in consensus (RQ 1), highlighting actors' ability to mix roles. Developers propose changes, full nodes adopt the changes to judge the validity of blocks and transactions, miners enforce consensus rules by generating blocks, and users can choose which consensus rules to abide by connecting to the full nodes in the network they prefer. The diagram includes simple payment verification (SPV), a method for nodes (e.g., miners and users) to perform partial verification without checking every consensus rule or storing the whole blockchain.}
  \label{fig:wenn}
\end{figure}

Blockchain nodes are complex and connect in an m-n fashion in the peer-to-peer network. The Wenn diagram in~\cref{fig:wenn} shows the complexity of the four roles in a blockchain and are abstractions from the capabilities actors have as network participants. The figure aims to show how actors or blockchain nodes come in many shapes, mixing capabilities and multiple roles. In contrast to related work~\cite{kiffer2017stick, de2016invisible, parkin2019senatorial, romiti2019deep, andersen2017patterns, kroll2013economics, 7176229, gencer2018decentralization}, we show that nodes, users, and miners are not concrete or exclusive roles. 

\Cref{fig:wenn} mainly distinguishes full nodes and simple payment verification (SPV) nodes. Full nodes validate based on all consensus rules and need all blockchain data. On the other hand, SPV nodes run partial consensus validation and connect to full nodes in the network to ensure the correctness of their state. It is also possible to have seed nodes to only relay blocks. However, it is not common or useful to only store and relay blocks without validation.

\begin{figure}[t]
    \centering
    \includegraphics[width=1\linewidth]{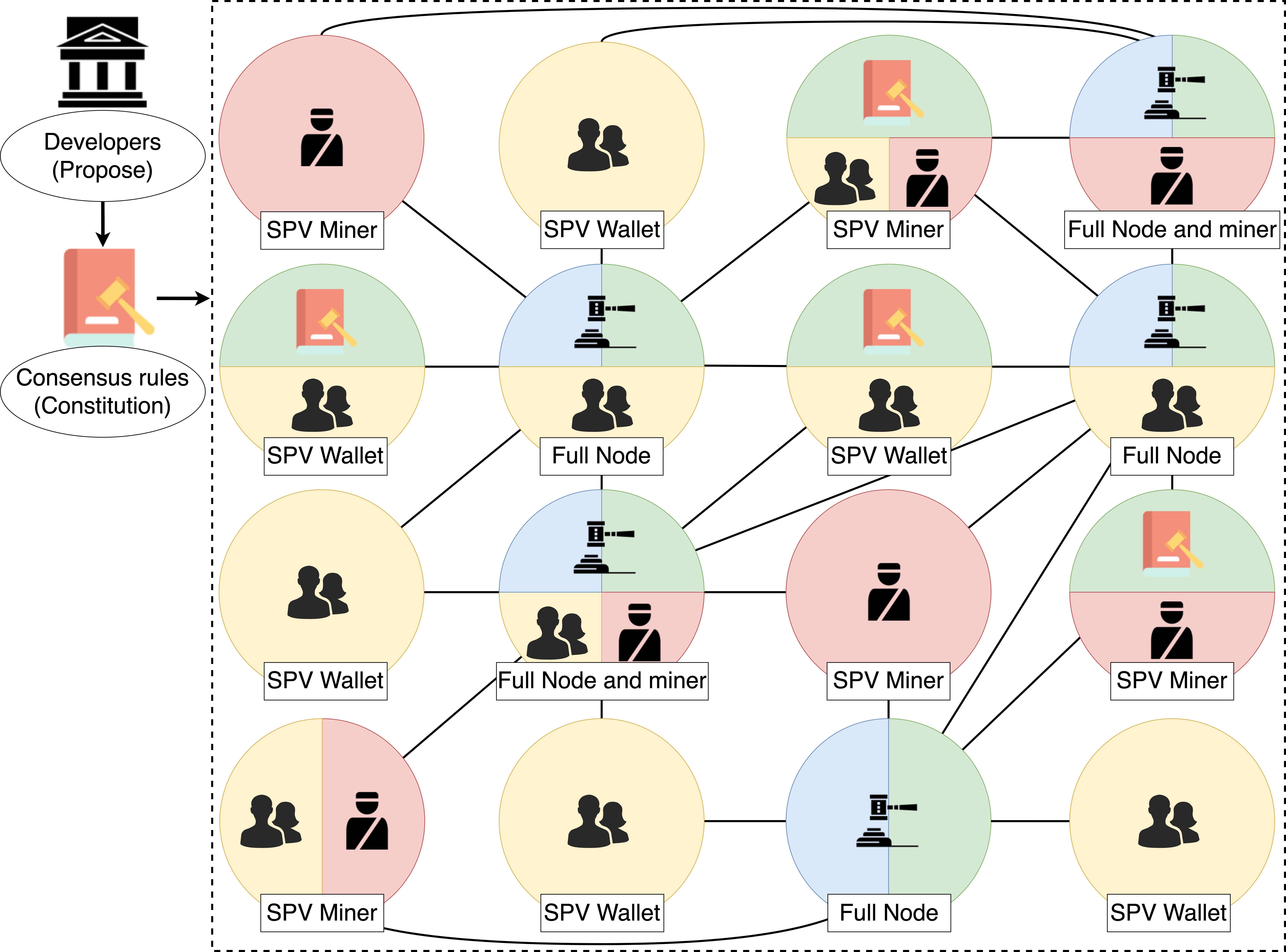}
    \caption{The diagram illustrates an example of a heterogeneous blockchain network where actors connect dynamically and combine different roles. Notice that all nodes connect to full nodes to synchronize with the network. The figure is an illustration and does not accurately resemble the network distribution in Bitcoin.
    }
    \label{fig:topology}
\end{figure}

We define the capabilities of the four abstracted roles in~\cref{fig:wenn} as follows: 1) Developers have access to all the components in the diagram and can \textit{propose} changes to the consensus rules (\textit{constitution}) without having to run a node. 2) Full nodes adopt consensus rules proposed by the developers and hold all blockchain data to enable full validation. By doing full validation, they \textit{judge} the correctness of proposed blocks from miners. 3) Miners run mining operations, and \textit{enforce} the consensus rules upon users by deciding to include transactions in blocks. Miners often do not contain all blockchain data nor validate all consensus rules. Instead, they perform simple payment verification (SPV) - a partial validation of transactions and blocks. 4) Users can be satisfied by running any wallet service. The users can \textit{choose} which wallet to use, which connects to certain full nodes - ultimately choosing the underlying \textit{constitution}. Users encompass everyone using a blockchain, including service providers such as wallets and exchanges.

\Cref{fig:topology} shows how a single node can represent one or many of the four roles simultaneously, illustrating how actors are irregularly connected and can have different and multiple roles.
Additionally, it demonstrates how consensus is maintained even though many nodes do not perform full validation and that nodes must eventually connect to a full node to ensure state integrity. Furthermore, full nodes interconnect to maintain consensus.

One can argue that the development of some repositories, such as Bitcoin Core, is centralised~\cite{parkin2019senatorial,de2016invisible}. However, the developers can only change that specific implementation to \textit{propose} consensus rules. They cannot force anyone to use them. The full nodes hold the consensus rules and \textit{judge} which chain their connected users will follow in case of a split. In contrast, miners may not validate all consensus rules to gain an advantage in the block-race~\cite{invalidBlocks}. In the end, miners decide the future of a blockchain, whether nodes will evolve in harmony or split into different blockchains.

Furthermore, miners \textit{enforce} the consensus rules upon users by including transactions and producing blocks. Without miners, the blockchain will stop extending. The miners can deploy a reduction fork without causing a chain split by having majority adoption or 100\% miner adoption in expanding and bilateral forks. There will always be an economic incentive to avoid chain splits because a consistent network will be perceived as more reliable. In contrast, an unstable cryptocurrency could depreciate. In the end, users can \textit{choose} which wallet to use, which full nodes to connect to, which network they will be part of, and, therefore, which consensus rules to follow. 

The four roles in Bitcoin differentiate the dynamics of blockchain from classic software development. Traditionally, there are mainly two roles: The developer and the user~\cite{guzzi2013communication, sharma2021extracting}. Users have less influence over the code and are limited to inquiries and feature requests~\cite{guzzi2013communication}. Furthermore, Sharma et al. highlight different developer roles: Benevolent dictator for life (BDFL), PEP author, PEP editor, and core developer~\cite{sharma2021extracting}. These distinctions might not matter in a permissionless blockchain where developers only make proposals while the network makes decisions.


\begin{mybox}
\textbf{Summary of RQ 1:}
\textit{Which roles and actors can influence consensus evolution governance?}
\begin{itemize}
  \item Developers \textit{propose} consensus rule changes.
  \item Full nodes \textit{judge} consensus rules by ensuring the validity of transactions and blocks.
  \item Miners \textit{enforce} the rules by choosing which transactions to include and generating blocks.
  \item Users \textit{choose} which full nodes they connect to, ultimately choosing which blockchain and corresponding consensus rules they like to use.
\end{itemize}
\end{mybox}



\subsection{RQ 2: The decision-making process of Bitcoin consensus evolution governance} 
\label{keyEvents}

\begin{figure*}[ht!]
  \centering
  \includegraphics[width=0.90\textwidth]{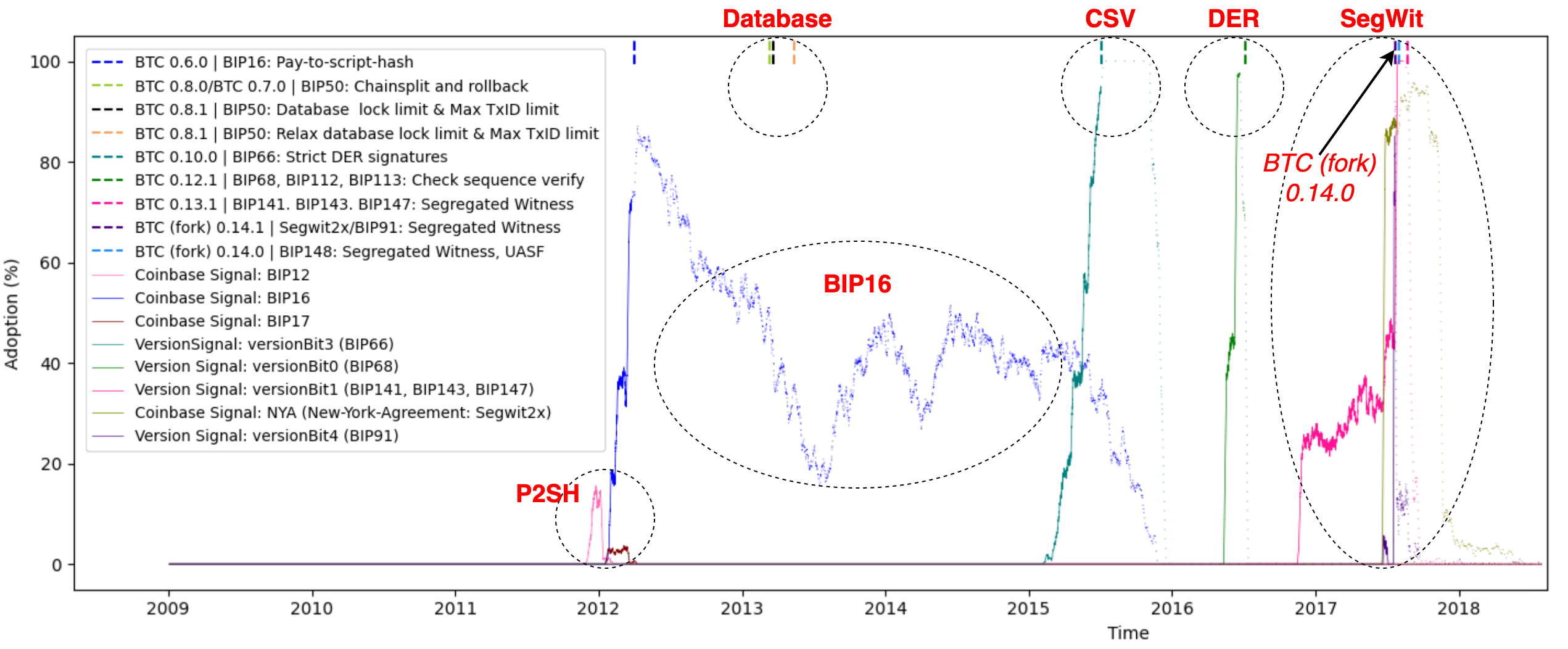}
  \caption{The graph illustrates successful key deployments in BTC. Solid lines show the miners attempting to coordinate and adopt new consensus rules by including signals in blocks. The changes were activated based on the adoption threshold, shown as dashed lines on the top. Signals are faded after activation to show that they no longer serve a purpose. 
  The included changes were mainly deployed using the miner-activated reduction fork (MARF) technique, except BTC 0.8.1 in \textbf{Database} and {~\textit{BTC (fork) 0.14.0}} which were user-activated, i.e., flag day activated forks without on-chain coordination.}
  \label{fig:BTCevolutionLight}
\end{figure*}

\begin{figure*}[ht!]

  \centering
  \includegraphics[width=0.90\textwidth]{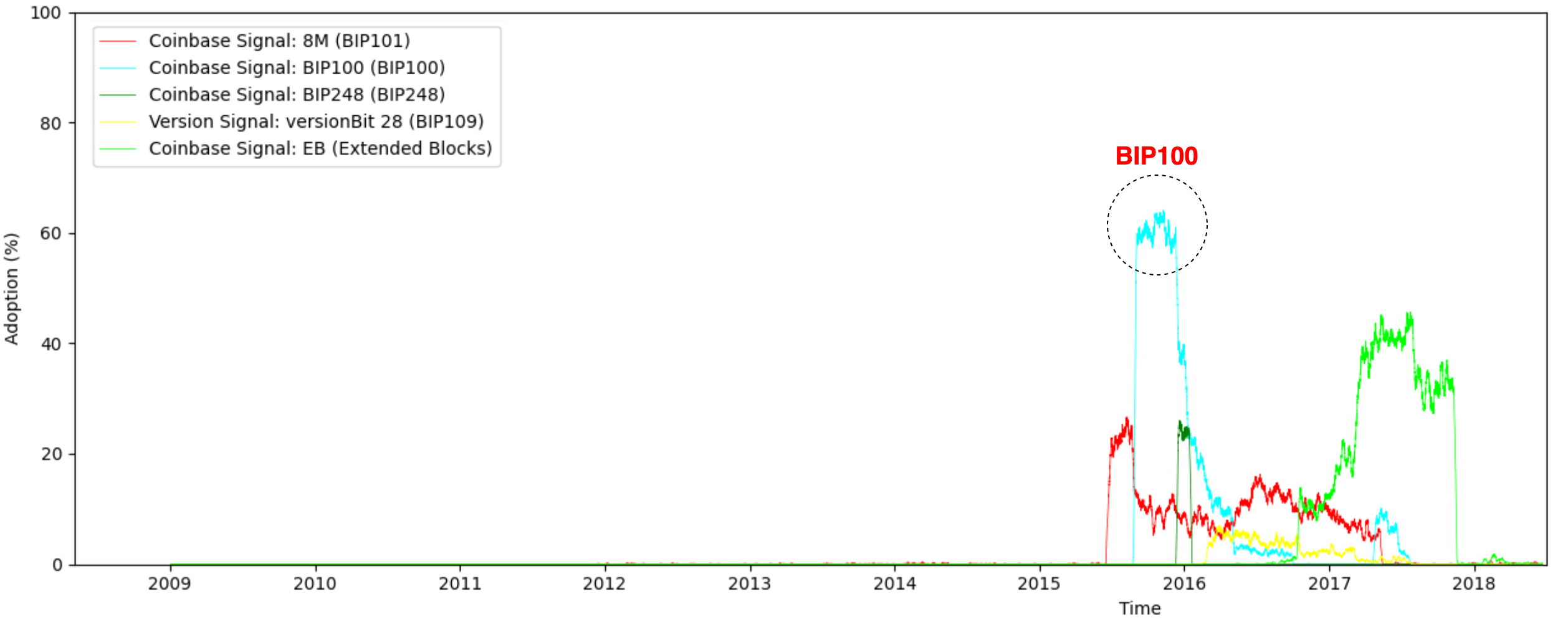}
  \caption{The graphs show failed attempts to coordinate the adoption of new consensus rules for increased blocksize. Using a miner-activated expansion fork (MAEF) technique, the miners included signals to show readiness. If the threshold was reached, those miners would have created a chain split. One of the attempts (\textbf{BIP100}) reached over 60\% adoption, which was close to the 75\% adoption threshold, resulting in a failed attempt.} 
  \label{fig:BCHevolutionLight}
\end{figure*}

In this section, we present key governance events to understand the decision-making process of Bitcoin consensus governance. \textbf{The key events} are identified by the labels in the text and~\cref{fig:BTCevolutionLight,,fig:BCHevolutionLight}, presenting on-chain signals and activation timings related to BTC's consensus evolution governance. The figures include key events until 2018, highlighting evolutionary decisions made in BTC. 
The y-axes show the percentage of signalling blocks based on the latest 1000 blocks, and the x-axes show the time. All the plots are retrieved from the BTC blockchain. At the same time, the graphs are separated to enhance readability. \cref{fig:BTCevolutionLight} contains events, on-chain signals, and activation timings related to deployed consensus changes. \cref{fig:BCHevolutionLight} contains plots of signals related to failed attempts to deploy blocksize increase. The graphs in \cref{fig:BTCevolutionLight} and \cref{fig:BCHevolutionLight} only include the deployment processes and events discussed in this section.
More detailed graphs are included in~\ref{signals} to show all on-chain activity, including Taproot, the latest consensus rule activation (2022).

~\textbf{P2SH consensus change.} During the nascent stages of Bitcoin, from late 2011 to early 2012, developers \textit{proposed} three conflicting reduction forks for pay-to-script-hash (P2SH) (BIP12, BIP16, and BIP17). Influential Bitcoin developers stood behind each proposal and proposed a miner-activated deployment strategy for each proposal. 

\begin{figure*}
  \includegraphics[width=1\linewidth]{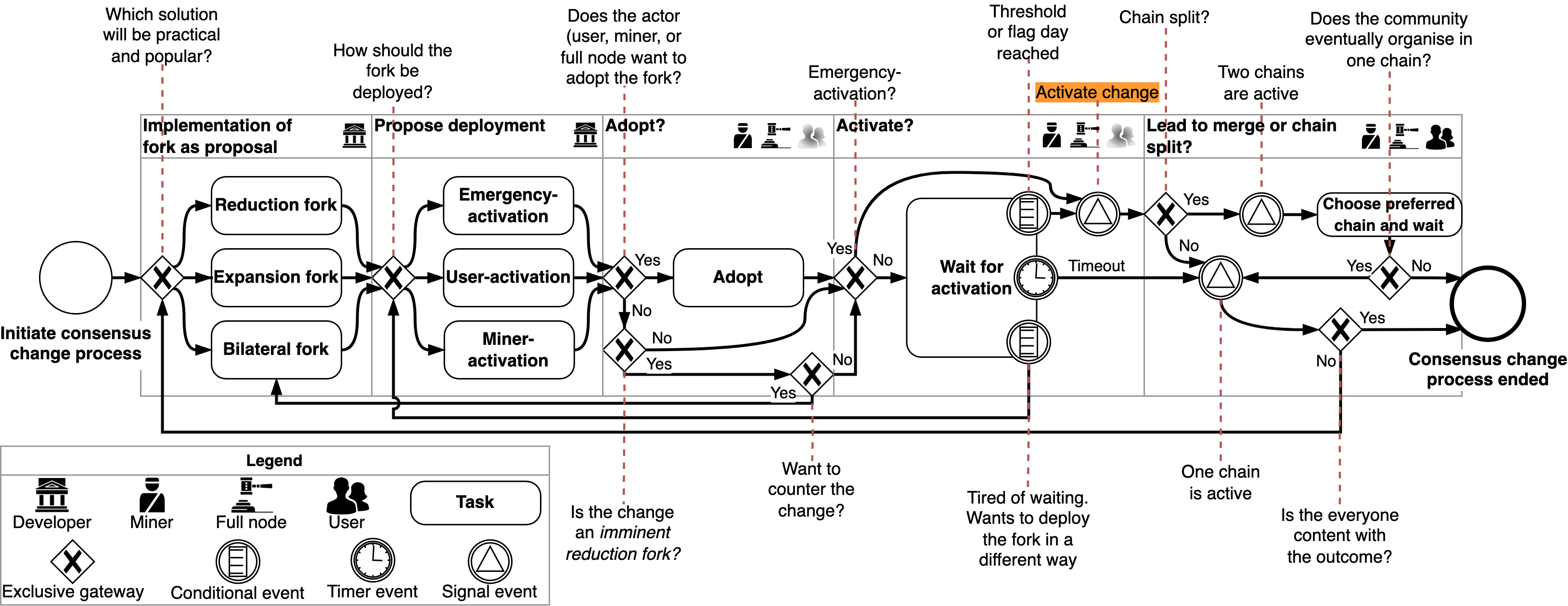}
  \caption{The {\tt BPMN} flowchart shows the process of evolving the consensus of a permissionless PoW blockchain like Bitcoin and the corresponding governance of the consensus change. The diagram is divided into five main stages, highlighting which roles can influence the outcome at each stage. In some cases, the user is greyed out because their influence is conditional.}
  \label{fig:decision-making}
\end{figure*}

The miners mainly adopted the BIP16 implementation, which would print ``/P2SH/" as a signal for readiness in the coinbase message in blocks. When 55\% of the last 1000 blocks contained the signal, the change would be considered ready to activate. BIP16 also contained a flag day to activate on 15th February 2012 based on the expected time to reach 55\% adoption. However, BIP16 lacked support by then, and the code was changed to activate on 1st April 2012 instead. This delay allowed the miners to reach over 55\% adoption. As shown in the last stage of~\cref{fig:decision-making}, the activation led to a merge of the consensus rule without a chain split because it was a reduction fork adopted by the majority of miners, and no conflicting forks were activated.

P2SH was the first coordinated upgrade on Bitcoin, activated through a primitive miner-activated deployment strategy known as is-super-majority (ISM). The strategy was primitive because the activation threshold was evaluated manually, and the timing was moved manually by making a new software release. On the other hand, P2SH was a significant advance in consensus evolution governance since Nakamoto implemented and included previous changes without discussion. Nakamoto's consensus changes were mainly activated on a flag day, similar to user-activated deployment. In contrast to P2SH, later deployments implemented a rolling window to activate automatically based on the threshold~\cite{self}. These observations highlight how the techniques for evolution governance have evolved to fit the needs of consensus change.


The BIP12 proposal was first discarded after discovering a bug allowing unlimited recursion. Afterwards, two conflicting proposals were created, BIP16 and BIP17, causing contention. As seen in~\cref{tycho}, the operator (Tycho) of Bitcoin's biggest mining pool (Deepbit) represented roughly 1/3 of the hash power and the hesitant opinion by being unwilling to support and \textit{enforce} the \textit{proposal} before it gained more traction and technical guarantees. 

\begin{lstlisting}[style=quote, caption={Tycho 2012-01-25 Forum, ID: 61125}, label=tycho]
I would like to add something about one of the reasons why I don't want to be the FIRST to adopt P2SH: I don't like doing beta tests in a production environment. In last 2 days I already got 3 messages from Gavin about new bugs found in his implementation: one "minor bug" and one "major bug" ("one critical line was dropped in a merge and missed in my testing"). Also some coins were possibly destroyed in the process because a bug caused block fees to be lost.
\end{lstlisting}

Eventually, with improvements, miner support, and influence from most of the Core developers, the miners activated BIP16. The contention of deploying P2SH and the manual change of the activation timing caused a delay in the deployment process, leading to some full nodes activating too early and causing connected miners to get stuck.

The consensus mechanisms of the blockchain restricted the developer's decisions on the implementation and deployment of P2SH. The political view was that the change should be compatible with old nodes. Therefore, the rule changes were implemented as a reduction fork through a ``hackish" solution to retain compatibility. Further, the deployment relied on majority adoption to avoid a chain split. The coinbase message was used as a governmental instrument to signal and measure adoption before activation.

The signal for BIP16 stayed active for over three years after the activation of P2SH. At this stage, the signal was not considered, although it stayed active because it was not removed from the code before 2015 \cite{6203}. This is visible after activation in~\cref{fig:BTCevolutionLight} where the blue line signalling BIP16 shows traces of miners turning off the flag manually as it goes down before going up again with major releases of Bitcoin Core, which still included the signal. 

In later versions from and including BTC 0.12.1 - ``CHECKSEQUENCEVERIFY" (\textbf{CSV consensus change}), the activation signals (i.e., versionbits) were configured to cease after serving their purpose. This behaviour can be seen in~\cref{fig:BTCevolutionLight} as an immediate drop in miner signalling after activation. The CSV deployment was the first performed by the modern miner-activation deployment strategy~\cite{BIP-0009}. The previous strategies (variants of ISM) had some flaws because they permanently reduced the set of allowed versionbits, could not handle parallel proposals, and did not allow permanent rejection of proposals~\cite{BIP-0009}.

~\textbf{Database consensus rollback}. The Bitcoin network experienced an accidental chain split caused by an unintended non-deterministic expansion fork when switching from Berkeley DB to Level DB in BTC 0.8.0 \cite{BIP-0050}. One could say that the full nodes were split into two alternative realities, following different chains, judged by different constitutions. The miners and users would follow the chain judged as valid by the full nodes they were running or connected to. As shown in~\cref{rollbackIRC}, developers (Luke-Jr, sipa, jgarzik, and gavinandresen) and pool operators (Eleuthria and doublec) coordinated an emergency-activated version rollback of Bitcoin to let all nodes reorganise to the legacy chain. This emergency-activated deployment shows a case where the centralisation tendencies in the developer team and pools streamlined the governance process.

\begin{lstlisting}[style=quote, caption={Developers: Luke-Jr, sipa, jgarzik, and gavinandresen. Pool operators: Eleuthria and doublec 2013-03-12 IRC: \#bitcoin-dev}, label=rollbackIRC]
<Luke-Jr> gavinandresen: sipa: jgarzik: can we get a consensus on recommendation for miners to downgrade? (...)
<Eleuthria> I can single handedly put 0.7 back to the majority hash power
<Eleuthria> I just need confirmation that thats what should be done (...)
<sipa> Eleuthria: imho, that is was you should do, but we should have consensus first (...)
<jgarzik> sipa, Eleuthria: ACK on preferring 0.7 chain, for the moment (...)
<gavinandresen> Eleuthria: if you can cleanly get us back on the 0.7 chain, ACK from here, too (...)
<Eleuthria> alright (...)
<doublec> ok, rolling back to 0.7.2
\end{lstlisting}

BTC's consensus mechanisms and technical landscape steered the community toward a rollback to the legacy implementation. Since BTC 0.8.0 was a (non-deterministic) expansion fork, the network would require all miners, full nodes, and users to upgrade to discard the legacy chain. On the other hand, a majority had to mine the legacy chain to discard the new chain. That is because all miners, full nodes, and users would accept the legacy chain when becoming the longest, per the longest chain rule.

Many users still relied on legacy implementations, such as Mt.Gox, the largest exchange at the time. If the community chose to proceed with BTC 0.8.0, it would be a long-lasting inconsistency with service outages. Only a rollback enforced by an emergency-activated reduction fork (EARF) could quickly gather the community on one chain. The chain split caused by 0.8.0 is shown as the first dashed line under \textbf{Database} in~\cref{fig:BTCevolutionLight}. 
The second dashed line under \textbf{Database} shows the emergency-activated temporary reduction fork (EARF) restricting the database lock limit and max TxID, making the Level DB implementation (0.8.1) compatible with the Berkeley DB implementations (\textless~0.8). The third dashed line is a flag day-based user-activated expansion fork (UAEF) included in 0.8.1 to fully enable Level DB and deprecate Berkeley DB implementations.

~\textbf{DER consensus change.} The deployment of BIP66 implemented in BTC 0.10.0 - "Strict DER signatures" demonstrates full nodes passing judgment on misbehaving miners. The quote in~\cref{SPVMining} emphasises the risk of mining without validation.

\begin{lstlisting}[style=quote, caption={Bitcoin.org \cite{invalidBlocks}}, label=SPVMining]
Note that the roughly 50% of the network that was SPV mining had explicitly indicated that they would enforce the BIP66 rules. By not doing so, several large miners have lost over $50,000 dollars worth of mining income so far. 
\end{lstlisting}

The incident occurred because miners falsely signalled readiness during the adoption stage, as seen in~\cref{fig:decision-making}. At the activation stage, it became apparent that about 50\% of the miners were not performing full validation according to new rules, causing competing forks of the chain. During the last stage, upgraded full nodes detected the consensus violations and rejected invalid blocks. Eventually, more miners and full nodes updated their consensus rule validation, and the new rules merged with the network.


~\textbf{SegWit consensus change.} The most contentious consensus rule change of Bitcoin and any blockchain so far is the deployment of SegWit in August 2017. During the adoption stage, the pink line in~\cref{fig:BTCevolutionLight} (year: 2017) shows the struggle to reach the required 95\% threshold, often referred to as an attempt to veto SegWit deployment. Many SegWit proponents were determined to force through the upgrade with a user-activated reduction fork (UARF) and set a flag day to exclude non-signalling nodes on 2017-08-01. They gained support from the SegWit2x community, who intended to enforce SegWit signalling triggered by an 80\% threshold. The SegWit2x community also intended to increase the blocksize of Bitcoin later, albeit without success. The SegWit2x community adopted BIP91 to reduce the evaluation window and reached its threshold on 2017-07-20. The original 95\% threshold was reached on 2017-08-08, and the planned miner-activated reduction fork (MARF) deployment technique then activated SegWit after a grace time of 2016 blocks on 2017-08-24. 

The deployment of SegWit highlights how external (SegWit2x/BIP91) and anonymous (BIP148) developers enabled miners to enforce the upgrade using a reduced threshold, flag day activation, and exclusion against the advice of developers behind the original proposal (BIP141, BIP143 and BIP147). The quotes in~\cref{sipaUASF,maxwellUASF} emphasise miners adopting a user-activated reduction fork (UARF - denoted as UASF or BIP148 in the quotes) even though it was discouraged by developers from the centralised core team and increased the risk of a chain split.

\begin{lstlisting}[style=quote, caption={Sipa 2017-05-17 Github: Pull 10417}, label=sipaUASF]
I believe there is widespread support for SegWit, and I sure as hell hope it activates at some point in time. Unsure if you're aware, but I'm the primary author of the SegWit proposal and patches. However, I don't think there is a rush, and BIP148 seems an unnecessarily risky way to bring that about.
\end{lstlisting}

\begin{lstlisting}[style=quote, caption={Gregory Maxwell 2017-04-05 Email: \#bitcoin-dev}, label=maxwellUASF]
I do not support the BIP148 UASF for some of the same reasons that I do support segwit:  Bitcoin is valuable in part because it has high security and stability, segwit was carefully designed to support and amplify that engineering integrity that people can count on now and into the future.
\end{lstlisting}

A minority introduced the UARF deployment technique as a governmental instrument to persuade the sceptics to adopt SegWit. The UARF was exclusive, meaning that patched nodes would discard blocks not signalling support for SegWit. Therefore, the proponents were willing to bet on Bitcoin's economic future to ensure SegWit adoption with two possible outcomes. First, a \textit{majority} of nodes discard non-signalling nodes, denying block rewards. Second, a \textit{minority} of nodes discard non-signalling blocks, making a chain split, causing the legacy network to lose a significant portion of its users and miners, leading to network disruption and economic instability. The more miners enforce signalling, the more it can persuade non-signalling miners.

The so-called ``big blockers" opposed SegWit, arguing it was an undesirable solution to scaling Bitcoin. The Bitcoin community debated solutions to increase the blocksize for years. The SegWit reduction fork became imminent, against the will of big blockers, who reacted by launching Bitcoin Cash. They released their proposal as a user-activated bilateral fork (UABF) to ensure incompatibility with the SegWit chain. Furthermore, user-activation ensured that the chain would take effect on the same day as BIP148, discriminating non-SegWit nodes. After the chain split and deployment of BCH, the remaining planned upgrades in BCH have all been deployed using flag day based user-activated expansion fork (UAEF) techniques.

There was a 12h:20m delay after the activation of BIP148 and before the activation of Bitcoin Cash. During this time window, all Bitcoin Cash node-produced blocks were excluded by a majority of miners. Consequently, the Bitcoin Cash-compliant miners wasted energy producing blocks the network would never accept. From this instance, we discern a fundamental disparity between consensus and classical software evolution. While forking is feasible in classical software evolution as well, the timing of its activation is not as crucial as in consensus evolution. During a majority-enforced~\textit{exclusive} fork~\cite{self}, opposing miners will collectively lose money for every discriminated block until becoming compliant or causing a chain split.

\cref{fig:BCHevolutionLight} shows several failed attempts to activate increased blocksize prior to SegWit with miner-activated expansion forks (MAEFs) between 2015 and 2018. 

~\textbf{BIP100 consensus change.} The signal with the highest support was BIP100, reaching support over 60\%. However, did not reach the threshold of 75\%. This expansion fork would have caused a chain split if adopted by less than 100\% of the miners. Furthermore, miners and users would follow one of the chains depending on which constitution they like, judged by the full nodes they chose to run or connect to. Additionally, at least one influential developer vouched for increased blocksize: Gavin Andresen, the successor of Satoshi Nakamoto. However, he did not convince a significant majority to adopt larger blocks. The quote in~\cref{ballsyAndresen} indicates Andresen's failed assumption of having decision-making power as a {\tt benevolent dictator}.

\begin{lstlisting}[style=quote, caption={Gavin Andresen 2015-04-18 \cite{ballsyAndresen}}, label=ballsyAndresen]
That may be what has to happen with the blocksize, frankly. I may just have to throw my weight around and say: This is the way it's gonna be, and if you don't like it, go find another project.
\end{lstlisting}

\subsubsection{Generalising the decision-making process}

Our findings from RQ2 give new insights into the decision-making process in consensus evolution governance. We depict the process in~\cref{fig:decision-making} using {\tt BPMN}~\cite{omg2011bpmn}, showing ~\textit{how governance and deployment intertwine.} 


~\textbf{Implementation of fork as proposal.} The process is initiated by identifying an issue. In this stage, developers present the implementation of a fork as a proposal. This approach contrasts with classic software evolution, where proposals undergo evaluation and decision-making before implementation. In that scenario, developers maintain authority over their codebase and determine whether to incorporate a proposal. Opposition developers can fork the codebase and introduce alternative proposals at their will.

Whether the proposed fork should be a reduction, expansion, or bilateral fork depends on the issue. A reduction fork is easier to deploy because it only requires a majority of miners to adopt the change. On the other hand, reduction forks may require workarounds to retain compatibility. An expansion fork does not consider compatibility restrictions and could enable more optimal solutions. Bilateral forks are rare and only seen during the chain split to launch Bitcoin Cash. A bilateral fork seems most viable to force a chain split during a minority fork, for instance, when countering an unwanted reduction fork.

~\textbf{Propose deployment.} Furthermore, developers propose conditions to activate the proposed change. There are three ways to activate a consensus change. Miner-activation is desirable to ensure activation based on a threshold proposed at this stage. On the other hand, user-activation launches on a proposed flag day. The last option, emergency-activation, is viable if there is an urgent flaw. For instance, BTC 0.16.3 and BCHN 0.18.2 were deployed using emergency-activation to address CVE-2018-17144 in Bitcoin and Bitcoin Cash. It was released as an emergency-activated reduction fork so the chains would split if the bug was exploited before majority adoption.

In most software development projects, developers offer only code without focusing much on the deployment, which is the concern of the users. On the other hand, in blockchain consensus evolution, the deployment proposal is a necessary integrated part of the consensus change. 


~\textbf{Adopt?} After developers have released a proposal, the rest of the network will decide whether to adopt it. Miners must adopt the new software to enable activation and, ideally, avoid a chain split. The proposal defines the exact conditions for deployment and activation.

Depending on the nature of a fork and its effect on the information flow in the peer-to-peer network communication, a certain amount of full nodes must adopt the change to avoid eclipsing~\cite{heilman2015eclipse}. For example, in SegWit, only upgraded full nodes relay signature information~\cite{self}. The risk of network eclipsing could apply to all fork types.

Users are not required to adopt a reduction fork. As a result, their icon in ~\cref{fig:decision-making} is made less visible by greying it out to indicate that their significance is conditional. However, users must adopt changes in wallets, exchanges, and other services for expansion and bilateral forks. 

Mining pool operators can adopt changes on behalf of miners. This could be done maliciously by including consensus changes without community approval. This could make the pools unpopular, and miners who disagree could change pool, mine independently, or take additional steps to counter a forceful consensus change.

For unwanted forks, one may take action to counter an ~\textit{imminent reduction fork}. Network participants must either take on the developer role or encourage other developers to make a new proposal. An imminent reduction fork can be countered by launching a bilateral fork as it was for Bitcoin Cash. However, suppose the unwanted fork is a minority-supported reduction fork, an expansion fork or a bilateral fork. In that case, countermeasures are unnecessary since these fork types cause the chain to split. 

The actors, especially miners, must consider the contention of a change to decide if it is economically reasonable. Contested changes could lead to contention and, in the worst case, a chain split. The fallout may further cause instability and depreciate the cryptocurrency, and it might be better not to change the consensus rules at all.

Full nodes and users must also position themselves in case of a possible chain split. Full nodes can adopt changes on behalf of connected users (including service providers such as exchanges). Users who disagree can connect to other full nodes. For instance, by using a different wallet.

In classical software evolution, adoption implies that changes will take effect immediately after deployment. However, this is not the case in blockchain because adoption is a slow process needing to wait for the conditions for activation, e.g., the adoption threshold or a set timing, before the consensus change activates. In permissionless PoW blockchains, such as Bitcoin, actors will first adopt the change individually in their own time and decision. The time while the network is adopting new software can be seen as an initialisation stage.  Therefore, ~\cref{fig:decision-making} includes a waiting period after adoption, before activation.

~\textbf{Activate?} Regardless of the individual user, miner, or full node's decision on adoption, it depends on the rest of the network and the conditions for activation, whether a partition or the whole network will commit to a change.
The different roles have the same impact in the activation stage as the adoption stage, resulting in the user icon still being greyed out at this stage in ~\cref{fig:decision-making}. 

Furthermore, the fork may or may not activate. Unlike traditional software changes, the activation here relies on coordination and co-decision of decentralised actors (miners, full nodes, and users). If the fork remains inactive, anyone can take on the developer role or encourage developers to return to the deployment stage and propose different conditions for activation. This happened for SegWit, where an anonymous actor released an implementation~\cite{BIP-0148} to discriminate miners that did not adopt SegWit. 

Emergency upgrades ignore the conventional processes of waiting for adoption and should be configured to force a chain split if necessary. Coordinated developer teams that quickly release a proposal and pools managing miners for quick adoption can enable swift activation.

~\textbf{Leads to chain split or merge?}
In the last stage, the new consensus rules may or may not be activated. If new consensus rules are activated, it could lead to a chain split or merge with the old rules in a single chain. In case of a chain split, the actors can choose which chain they prefer. In time, the community might again organise into a single chain and abandon one of the forked chains. On the other hand, if the rules merge into a single chain, they can only be reverted by making a new proposal and restarting the consensus change process. In classical software evolution, the participants commit to an implementation before deployment and stick with that. This differs from the dynamics of blockchain evolution. In case of a chain split, one could choose which deployed instance to adopt after launch.

In theory, blockchain community members may create chain splits and choose whichever chain they want. On the other hand, there is no utility in a new chain if the actors act alone or in a small group. For instance, anyone can download the Bitcoin implementation and adjust a few consensus rules to cause a chain split. However, if they are lonely participants on their chain, they are essentially running a small experiment - a test net with no value. Our study asserts that anyone can initiate changes in blockchain and participate in any network. Regardless, no one wants to run a blockchain network alone, and high participation and decentralisation prosper a healthy and valuable blockchain. Likewise, attempts to force changes upon a blockchain network require caution, as it may jeopardise the harmony that has taken years to establish.

\begin{mybox}
  \textbf{Summary of RQ 2:}
  \textit{What is the decision-making process for consensus changes?}
  \begin{itemize}
    \item The consensus change decision process starts from the developers' proposal, including the fork type, implementation, and deployment strategy with conditions for activation. Miners, full nodes, and users decide when and how to adopt the changes. Activation is done automatically without human intervention, either by a flag day (user-activation), when reaching the threshold for compliant miners (miner-activation), or immediately when adopted (emergency-activation). Then, miners, full nodes and users decide which chain to follow if there are chain splits.
    \item The proposal adoption and activation could be a slow decentralised community coordination to adopt until they automatically activate upon meeting the conditions. The decision to adopt is largely influenced by the economic considerations of the miners and users.
    \item The emergency upgrades enable swift activation by ignoring conventional processes and force a chain split if necessary. Centralisation tendencies in, developer teams and mining pools are useful in emergencies. The centralised Bitcoin Core group can quickly release a patch, and a few mining pools can quickly adopt the changes for a majority of the miners.
    \item Compared to classical software evolution, the key difference in blockchain evolution is that:
    \begin{itemize}
        \item It is not viable to decide whether to include a change based on only the rationale. The implementation must be provided for a proposal to be considered for inclusion.
        \item Development and deployment must be proposed together.
        \item Just because someone decides to adopt and commit to change and consensus evolution, it does not guarantee activation.
        \item The blocks and rewards of miners could be discriminated against and omitted because they are slow to adopt changes even though they produce otherwise compatible blocks.
    \end{itemize}
  \end{itemize}
\end{mybox}

\subsection{RQ 3: Roles and decision-making power distribution}

\subsubsection{The impact of contested change and fork type on consensus change governance}
\label{governance}

The analytical codes related to the social/political category (\cref{fig:codesCategories}) depict how the consensus change governance process relies on the level of arguments surrounding the change. A change without contest requires low effort in convincing the community and scheduling deployment. Centralisation tendencies in development and mining operations can streamline the deployment of uncontested changes. Changes that can be deployed most quickly are related to fixing an immediate issue in the blockchain. In that case, the developers are in a hurry to implement a solution while miners and mining pool operators eagerly install it.
The fix of CVE-2018-17144 is the latest example of developers and pools streamlining deployment (versions: BTC 0.16.3 and BCHN 0.18.2), where the Slush pool collaborated with developers to deploy the upgrade before they released the code the day after discovery~\cite{CVE-2018-17144}. 

However, for planned changes (i.e., anything other than emergency-activated changes), the deployment must happen publicly and with enough time to review the proposal and implementation. Back-room-like agreements between pools and developers will cause havoc in the community. The closest example of such behaviour was observed during the Hong Kong Roundtable Consensus~\cite{hkrtc} and New York Agreement \cite{nya}, where prominent mining actors and central developers agreed to pursue a 2MB expansion fork on top of SegWit. The proposal was named Segwit2x, and the proponents held around 80\% of the network's hashing power. However, the proposal did not gain enough traction in the community and ended up being cancelled~\cite{segwit2x}.

Whether a change is a reduction, an expansion, or a bilateral fork is vital because reduction forks can be deployed with only a majority of miners. In contrast, expanding and bilateral forks require 100\% support from miners to prevent a chain split. Additionally, services like wallets and exchanges might require changes to support expanding and bilateral forks. Therefore, expanding and bilateral forks rely more on support from users to adopt the new chain, maintain its value, and pay fees. 

Theoretically, a majority of miners may bypass the wider community and force activation of a contentious proposal. However, without users' support, the cryptocurrency could depreciate. Therefore, the miners depend on user adoption to financially support the system, as much as users depend on miners for the system's security. These aspects highlight the decision-making power of users, which is also emphasised in~\cref{drivechain}.





\begin{lstlisting}[style=quote, caption={2023-10-13 Peter Todd and Dave Weisberger \cite{drivechain}}, label=drivechain]
<Dave Weisberger> (Hypothetical) I'm going directly to the miners, and I'm making the case for why they need to adopt drivechain, and I'm selling them on the benefits of it, and frankly, I don't care what anyone else's views are because it is the miners at the end of the day that I have to convince. If the community has a problem with that, then what can they do to stop them?
<Peter Todd> What you are telling miners here is that this is going to be a shitshow, and if you go through this and add all this uncertainty to Bitcoin. Ultimately, you are probably either going to just reduce the price in general, or you will go have a UASF to undo drivechains and that is a huge mess. (...) It is a much bigger mess than just doing nothing. (...)
\end{lstlisting}

Regardless of the fork type, contested changes prolong the consensus change process and could lead to contention. Furthermore, contested changes may become a source of controversy where irrational voices express FUD (fear, uncertainty, and death) or shill to express the advantage and fortune that can be made through the change. In case of contention, developers should use the deployment techniques responsibly, optimally leading to deployment without a chain split or no deployment at all. Whether miners would support a contented proposal relies on seeing the benefit of that change and whether the users are willing to adopt and pay for the system's services afterwards.

\subsubsection{Decision-making power}

The results from RQ 1 established the role definitions for blockchain. In this part, we focus on how they can use their role to influence the process derived from RQ2. 

\begin{figure}
  \centering
  \includegraphics[width=1\linewidth]{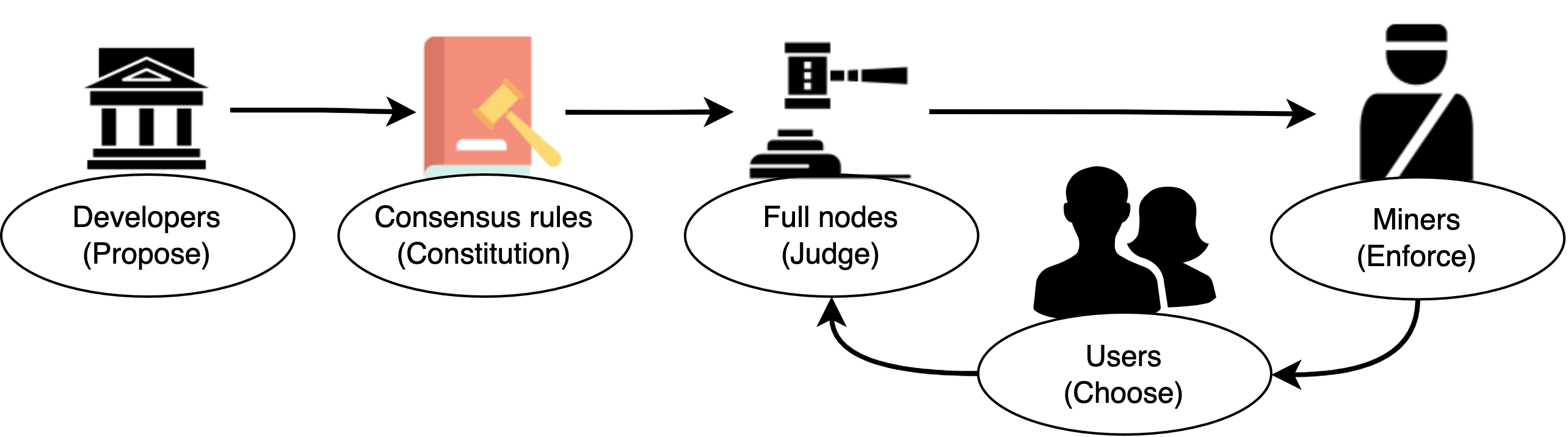}
  \caption{Blockchain governance: roles and decision-making power distribution. The pattern closely resembles the tripartite system's division and balance of power~\cite{tripartite}. Except that anyone can participate in any and as many roles as they want. The miners can enable consensus change. However, they are incentivised to do so only with significant support from the community, including other roles.}
  \label{fig:power}
\end{figure}

We argue that blockchains hold their value in participation and adoption, and all the actors will generally benefit from collaborating. Whether a rule change is worth deploying is a trade-off between the impact of the new functionality and the impact on the network's consistency~\cite{self}. The balance of power for the roles deciding the evolutionary path for a decentralised distributed ledger is illustrated in~\cref{fig:power}, which demonstrates a flat and decentralised decision-making power structure among miners within the blockchain. As highlighted in~\cref{fig:power}, anyone can act as a developer and write code, but \textbf{it ultimately falls upon the miners to adopt consensus changes}. Bitcoin operates neither as an oligarchy nor a democracy; rather, it functions through consensus, driven by miners and governed by the longest valid chain rule. \textbf{The concept of validity remains subjective, allowing each individual to choose their preferred consensus implementation.} However, consistency is crucial, and a cryptocurrency's value hinges on its usage and adoption. Additionally, security in Proof of Work (PoW) blockchains relies on the total hash power and its distribution across the network. Therefore, avoiding controversial changes or reconciling with the majority fork might be desirable in case of a chain split. 




A limitation of Bitcoin's consensus evolution governance structure is that few quality assurance indicators allow miners, full nodes, and users to assess the security guarantees of a change proposed by developers. Therefore, insufficient knowledge or expertise of the changes compromises their power to make correct decisions. Although such processes exist, they are often not well communicated. For instance,~\cref{ACK} shows a reviewer's approval of the code for a reduction fork (BIP66~\cite{BIP-0066}). In total, there are six participants with ten comments on the proposal. Anyone who is not an expert in cryptography would have to trust the statements of these participants in this thread without more insights into the extent of assurances tied to the BIP66 consensus change. 
To improve the situation, we suggest that developers should provide more indicators before initiating consensus changes, such as proof of audit and testing new code. One might argue that the level of auditing and testing is sufficient. Regardless, developers must communicate these processes well to allow the community to make qualified decisions.

\begin{lstlisting}[style=quote, caption={Laanwj 2015-01-28 Github: Pull 5713}, label=ACK]
tested ACK. Verified that the deployment behavior as mentioned in BIP66 is indeed what is enforced.
\end{lstlisting}

\begin{mybox}
  \textbf{Summary of RQ 3:} \textit{How is decision-making power distributed among the different roles in consensus evolution?}
    \begin{itemize}
      \item There is not a single entity or role that singlehandedly holds the power to change consensus. Developers, miners, full nodes and users depend on each other for a blockchain to serve its purpose and retain its value. 
      \item A divided community cause economic instability. Miners can force through consensus changes. However, if they cause chaos and depreciation, then miners are better off avoiding them. These economic aspects give users decision-making power.
      \item Consensus changes often deal with intricate mechanisms, and the implications and risks of a change are often merely understood by a handful of developers. Miners, full nodes and users often trust the developers and adopt their changes without sufficient quality indicators and security guarantees.
    \end{itemize}
\end{mybox}

\section{Discussion} 
\label{discussion}

\subsection{Comparison with related work} 
\label{relatedWork}

Filippi et al.~\cite{de2016invisible} and Parkin~\cite{parkin2019senatorial} focus on governance in Bitcoin, specifically on the inherent power structures. Bitcoin is described as a technocratic system relying on exceptionally skilled developers' knowledge, a pattern we also recognise within the Bitcoin Core repository. Different from Filippi et al.~\cite{de2016invisible} and Parkin~\cite{parkin2019senatorial}, our results highlight the insignificance of centralisation patterns for contested changes. As observed for BIP-148, the Segwit deployment was forced through by an external and anonymous developer, even though discouraged by Bitcoin Core developers. 
Our view is that the Bitcoin Core group do not control Bitcoin evolution. However, it is a challenge that other stakeholders might not be capable of judging and making an informed decision and rather trust the developers.


Furthermore, Filippi et al.~\cite{de2016invisible} and Parkin~\cite{parkin2019senatorial} isolate governance conducted by the infrastructure (i.e., the blockchain) and the architects of the infrastructure, whereas our analysis deems these aspects highly intertwined.
For instance, interdependence clearly shows when the fork type determines whether a majority can deploy a change without a chain split, which is eligible for reduction forks. In contrast, expanding and bilateral forks must be deployed with 100\% miner support to avoid a chain split. Alternatively, one must consider whether deploying with a chain split is economically reasonable given the potential disruption and whether a users will use the new chain. 




Andresen and Bogusz ~\cite{andersen2017patterns} address Bitcoin governance from the viewpoint of "how and when code development practices combine into a pattern of self-organizing." They demonstrate how and when different forks occur, being both consensus forks and forks in community groups.
Their work does not emphasise how consensus changes are realised. In contrast, our study solely focuses on consensus changes and shows the roles of evolution governance, the decision-making process, and power structures.

Kroll et al.~\cite{kroll2013economics} argue that Bitcoin needs governance more than the participants might be willing to admit and that tendencies of governance structures are evolving, primarily as a collaboration between pool operators and developers. They are supported by the database example from the 2013 consensus failure and rollback~\cite{BIP-0050}, where developers and pool operators collaborated to make an exception from the longest chain rule. Centralisation tendencies in pools and developers may streamline the deployment of an upgrade.
However, we argue that they cannot dictate miners, as miners can change between pools if they disagree on the direction to steer Bitcoin.
We also show that in cases of contested changes, the centralised entities no longer have control over the consensus evolution governance. Furthermore, our findings emphasise new aspects of how governance and deployment intertwine in blockchain, which is an essential difference between classical software  and blockchain evolution.

The articles~\cite{7176229,romiti2019deep} consider pool centralisation and underline the risk of majority attack and block withholding attack by colluding pools. We recognise these aspects but argue that these threats do not give pools control to perform persistent consensus changes. 

Gencer et al. ~\cite{gencer2018decentralization} argue that no empirical support suggests that miners check pools' good behaviour and that detection of majority attacks might be impossible. In contrast, in terms of consensus change, our findings show that pools have been called out based on their behaviour because they are reluctant to support consensus changes during the deployment of P2SH~\cite{boycott}, support unpopular consensus changes, such as SegWit2x~\cite{hkrtc,nya}, or cause consensus failure, e.g., during the activation of BIP66~\cite{invalidBlocks}. 

Gencer et al.~\cite{gencer2018decentralization} argued that the majority-attacks are hard to detect. Regarding malicious consensus governance inference, our results show that it is easy to determine if pools perform malicious consensus evolution because that would be visible as code changes, flag day, or signals in the blockchain. Therefore, pools might be abandoned by their miners if they support controversial consensus changes.


%



Other blockchains show tendencies of self-regulation and governance. MakerDAO~\cite{makerdao} implements an emergency shutdown, which can be initiated by token holders. A re-deployment may require human intervention and a vote among the token holders. Similarly, Dash~\cite{sporks} implements a mechanism for network-wide rollback initiated by the developers. These mechanisms can be compared to the Ethereum DAO hack~\cite{kiffer2017stick} where an emergency was resolved through a voting period followed by a rollback. Our results show that Bitcoin deployment mechanisms are also evolving and have utilised similar emergency-activation strategies. Typical scenarios are the spontaneous reorganisations performed due to the faulty implementation in BTC $<$ 0.3.10 and the deployment of the incompatible BTC 0.8.0. The reason and policy for deploying the rollbacks in these blockchains may differ, but the deployment strategies are similarly realised through emergency-activation.


The main similarities between blockchain evolution and open software evolution can be seen in the decision-making on a code repository level~\cite{guzzi2013communication, sharma2021extracting}. Like the open source evolution, anyone can propose blockchain changes, depending on whether the repository's maintainers want to include that proposal. Blockchain presents a unique aspect in which the deployment decision is not only based on the maintainers' preference but also the opinion of the larger community: Miners, full nodes, users, and possibly developers of other implementations of the same protocol. Even when a change is included in a repository, it does not mean it will be adopted and activated in the network, because the change requires network consensus to be realised and activated after being accepted in the repository.

\subsection{Implications}
\label{implications}


Our research indicates that, in the blockchain-based system, the strength of developers' influence on consensus changes mainly lies in confidence in their proposals. Pool operators can help coordinate to resolve a consensus failure. The results of RQ3 illustrate that any ill-intended consensus change has many restrictions. It can only be effective for short-term attacks to censor blocks and transactions through a reduction fork and could lose traction unless accepted by miners and users over time. 

We demonstrate the decentralisation of governance processes for consensus change and reassure blockchain practitioners and those worried about centralised governance of consensus rules. However, a challenge is that the decision-makers in Bitcoin (miners, full nodes, and users) often lack insights in terms of quality indicators for proposed changes proposed. We suggest they demand more insights, for instance, by more visible testing and review.

Considering governance processes and decision-making power distribution among roles, recent events have shown the U.S. government attempting to govern Ethereum by censoring Tornado Cash transactions - an Ethereum mixer linked to cybercrimes~\cite{ofac}.  
However, it was not further investigated because this is happening in Ethereum. Hypothetically, censorship could be proposed as a reduction fork. That would be effective for censorship, although the network would probably not accept it. In Ethereum, many validators avoid including sanctioned transactions in blocks as a policy and increase the confirmation time for those assets. Alternatively, in Bitcoin, miners could utilise "feather-forking"~\cite{featherfork} to avoid mining on top of incompliant blocks in attempt to create block collisions. Feather-forking also discourages other miners from mining sanctioned assets because they might lose their block reward. This type of persuasive governance might not be practical in Ethereum because attempts to revert blocks could cause penalties~\cite{slashing}. In Bitcoin, this would not be well received because it conflicts with Bitcoin's liberal vision. If adopted, it could cause contention and depreciation.

\subsection{From benevolent dictator to decentralised governance}
\label{benevolent}

The findings of RQ2 illustrate how the evolution processes adapt to accommodate the requirements of consensus change.
Bitcoin was established as a system operating without trusted authorities~\cite{nakamoto2019bitcoin}. 
However, Nakamoto did not mention how to govern the maintenance and evolution of Bitcoin. The indicators are the longest chain rule, and that {\tt version 0.1} sets the core design in stone for the rest of its lifetime~\cite{nakamoto2019bitcoin}. Ironically, Nakamoto more or less acted as the trusted authority or {\tt benevolent dictator} in Bitcoin until his absence. During that time, he patched consensus-related code six times (0.1.6, 0.3.5, 0.3.6, 0.3.7, 0.3.10, and 0.3.12), mainly through emergency-activation or user-activation by block height timing. 
Nakamoto also created an emergency alert system to notify nodes of urgent issues. Nakamoto can be seen as the trusted authority of Bitcoin because the community probably would have accepted any sane code change, in contrast to the current mature and more conservative community.

Furthermore, Nakamoto left Bitcoin after staying for less than two years. There are many speculations about why Nakamoto left the project at that given time. Our analysis indicates that he planned to step down eventually when he no longer needed to maintain the project. Nakamoto did not add any new utility to Bitcoin after release, and the patches made before his absence were mainly bug fixes. Additionally, the community had grown with developers, full nodes, miners, and users to prosper further adoption.

The latest consensus change by Nakamoto indicates why it was time to step down as a benevolent dictator. That was in version 0.3.12, where the block size limit was enforced at 1MB. The change was created without traces of discussion. People started questioning the change within a month of its release before becoming the most controversial topic in Bitcoin. The block size debate could have looked different if there had been consensus on the initial limit, its reasoning, and whether to change it later. Miners no longer accept changes naively since they actively participate in the BIP process and deployment.










\subsection{Threats to validity}

We investigate governance in decentralised blockchains based on Bitcoin and Bitcoin forks. Bitcoin is a one-of-a-kind system and the longest-living blockchain project. Our results can be generalised to other decentralised software systems. 
However, the results apply more confidently to systems with a similar network structure and consensus model. Other decentralised constructs, such as layer 2 (Lightning Network~\cite{LN}), or layer 0 (Polkadot~\cite{polkadot}, and Cosmos~\cite{cosmos}) could be comparable as they mature.

Regarding internal validity, we emphasise the repeatability and verification of our research by thoroughly describing developing theories based on the available data. The data used is public and available through the archives and references in our paper. The samples~\cite{exportedSamples} used for analysis are also provided as exported from {\tt Atlas.ti}. Furthermore, we show nuanced perspectives with results derived from data triangulation and snowballing.
The sheer amount of data and the few researchers dedicated to this project may raise concerns about missing data or analysis. To address this, the three authors thoroughly evaluated the results through regular cross-author discussions. The recursive process in grounded theory made us look closely at BCH even though the initial data was gathered from BTC. BCH is interesting because the community has a different culture regarding consensus changes than BTC, where functionality might trump consistency. 


Regarding external validity,  we also compare the case of Bitcoin with other public blockchains to ensure generalisability. The research method included theoretical saturation, triangulation, and snowball sampling, which led to insights from other systems, such as MakerDAO and Dash, mentioned in~\cref{relatedWork}. Additionally, a research assistant investigated consensus changes in Ethereum 1.0. These findings did not expand the theory of our work. However, aspects like the U.S government's involvement in Ethereum 2.0 were considered in~\cref{implications}.


\section{Conclusion} 
\label{conclusion}

Consensus rule evolution is challenging and could cause suspended services, lost mining revenue, double-spending, and replay attacks. Furthermore, centralisation tendencies among developers and miners raise concerns about the consensus evolution governance. We address these concerns by focusing on Bitcoin to study evolution governance in permissionless PoW blockchains. The research was conducted using grounded theory, gathering evidence from 34 consensus forks, including off-chain governance and on-chain traces of evolution governance.

In contrast to concerns about centralised tendencies in pools and repositories, we show that consensus changes in Bitcoin fall back on the individual miners and the economic influence of users. Pools that act against the system's interest can only cause short-term harm and could quickly be abandoned by miners. Repository maintainers are in a position to propose changes but rely on community support. This paper provides empirical evidence that the decentralised consensus mechanisms in blockchain restrict the centralisation tendencies observed for developers and pools. Thus, the decision-making power to make permanent changes in consensus rules lies with the miners, constructing the longest chain, restricted by the different deployment techniques and dependence on user adoption. 

Throughout this study, we show which roles participate in blockchain governance and how they influence changes. We argue that the centralisation tendencies are not as significant as they seem and that developers and pools have limited decision-making power. However, we suggest that miners, full nodes, and users exercise their power more and demand quality assurance indicators to avoid accepting developers' changes depending on their ethos.

Furthermore, we show novel observations differentiating blockchain evolution governance from classic software: 1) Blockchain proposals must be completely implemented before they can be considered for adoption. 2) The implementation of a change and its deployment must be proposed together. 3) Adoption of a proposal does not necessarily lead to successful deployment through activation. 4) Under certain circumstances (exclusive forks), opposing participants (miners) could get punished until they become compliant or cause a chain split.

For future research, a deeper analysis of hash power distribution among individual miners could provide invaluable insights into the potential for collusion and its impact on blockchain evolution governance. Understanding the factors influencing miners' decisions to switch between pools and how this relates to consensus governance can reveal the extent of miners' active participation in the governance process. Additionally, examining the centralisation of economic resources among users can highlight how a few powerful actors could potentially sway the blockchain's direction. These directions not only build upon our current findings but also open new avenues for understanding the complex dynamics of blockchain governance.



\section{Acknowledgements}
Thanks to Peter Halland Haro and Peter Oyinloye Damilare for their help reviewing the article before submission.

\section{Funding}
This work is jointly supported by the National Key Research
and Development Program of China (No.2019YFE0105500)
and the Research Council of Norway (No.309494).



\bibliographystyle{elsarticle-num} 
\bibliography{elsarticle-template-num}

\appendix
\section{Data collection} \label{dataCollection}


\begin{table*}[]
    \footnotesize
    \setlength{\tabcolsep}{2pt} 
    \centering
    \caption{Overview of consensus changes in Bitcoin Core and applied search strings.} 
    \begin{tabular}{llllll}
    \textbf{Version} & \textbf{Consensus change} & \textbf{Maintenance type} & \textbf{Search strings} \\ 
    \hline
    BTC 0.1.6 & Time based locking nLockTime & Corrective & nLockTime \\
    \hline
    BTC 0.3.5 & CVE-2010-5137 \& CVE-2010-5141 & Corrective & CVE-2010-5137, CVE-2010-5141, 0.3.5 \\%
    \hline
    BTC 0.3.6 & Disable/enable opcodes & Adaptive \& preventive & OP\_CHECKSIG, OP\_NOP, 0.3.6 \\ 
    \hline
    BTC 0.3.7 & Separate scriptSig \& scriptPubKey & Corrective & scriptSig, scriptPubKey, 0.3.7 \\%
    \hline
    BTC 0.3.10 & Output-value-overflow CVE-2010-5139 & Corrective & overflow bug, 184, CVE-2010-5139, 0.3.10, 74638, 74 638 \\ 
    \hline
    BTC 0.3.12 & 20 000-signature operation limit \& 1MB blocksize & Corrective & MAX\_BLOCK\_SIZE, 79400, 79 400, \\
    & & & MAX\_BLOCK\_SIGOPS, 0.3.12 \\
    \hline
    BTC 0.6.0 & BIP30: Duplicate transactions CVE-2012-1909 & Corrective & bip30, bip 30, duplicate transactions \\
    \hline

    BTC 0.6.0 & BIP16: Pay-to-script-hash & Perfective & bip12, bip 12, bip16, bip 16, bip17, bip 17, bip18, bip 18, \\ 
    & & &  OP\_EVAL, P2SH, pay-to-script-hash, 0.6, 0.6.0 \\
    \hline
    BTC 0.7.0 & BIP34: Include block height in coinbase & Corrective & bip34, bip 34, 0.7, 0.7.0, 227835, 227 835\\
     & & & \\
    \hline
    BTC 0.8.0 & BIP50: Migrate from Berkeley DB to LevelDB & Corrective \& perfective & bip50, bip 50, leveldb, berkeley db, 225430, 225 430, \\ 
    & & & max 4 500, reduce blocksize, lock limit, 0.8 \\
    \hline
    BTC 0.8.1 & BIP50: Database lock limit \& Max TxID limit & Corrective & -||- \\ 
    \hline
    BTC 0.8.1 & BIP50: Relax database lock limit \& max TxID limit & Corrective & -||- \\ 
    \hline
    BTC 0.9.2 & BIP42: 21 million supply & Preventive & bip42, bip 42, 21 million, 21 000 000, 0.9.0, \\ 
    & & & 13,440,000, 13440000, 13 440 000 \\
    \hline
    BTC 0.10.0 & BIP66: Strict DER signature & Corrective & bip66, bip 66, strict der signatures, 0.10.0, 0.9.5 \\
    \hline
    BTC 0.11.0 & BIP65: Check lock time verify & Perfective & bip65, bip 65, check lock-time verify, CLTV, \\ 
    & & & CHECKLOCKTIMEVERIFY, 0.10.4, 0.11.2 \\
    \hline
    BTC 0.12.1 & BIP68, BIP112, BIP113: Check sequence verify & Perfective & bip68, bip 68, bip112, bip 112, bip113, bip 113, \\ 
    & & & Check sequence verify, Relative lock-time, \\
    & & & CHECKSEQUENCEVERIFY, CSV, 0.11.3, 0.12.1 \\
    \hline

    BTC 0.13.1 & BIP141, BIP143, BIP147: Segregated Witness & Perfective, corrective, & bip141, bip 141, bip143, bip 143, bip147, bip 147, \\ 
    & & adaptive \& preventive & segwit, 0.13, bip91, bip 91, bip148, bip 148, \\
    & & & segregated witness, 481824, 481 824 \\
    \hline
    BTC 0.14.0 & CVE-2018-17144 & Perfective &  -\\
    \hline
    BTC 0.16.3 & Fix CVE-2018-17144 & Preventive & -\\
    \hline
    BTC 0.21.1 & BIP341, BIP342, BIP343: Taproot & Perfective \& adaptive & -\\
    \hline

    \end{tabular}

    \label{tab:changes}
\end{table*}

\begin{figure*}[t]
  \centering
  \includegraphics[width=0.9\linewidth]{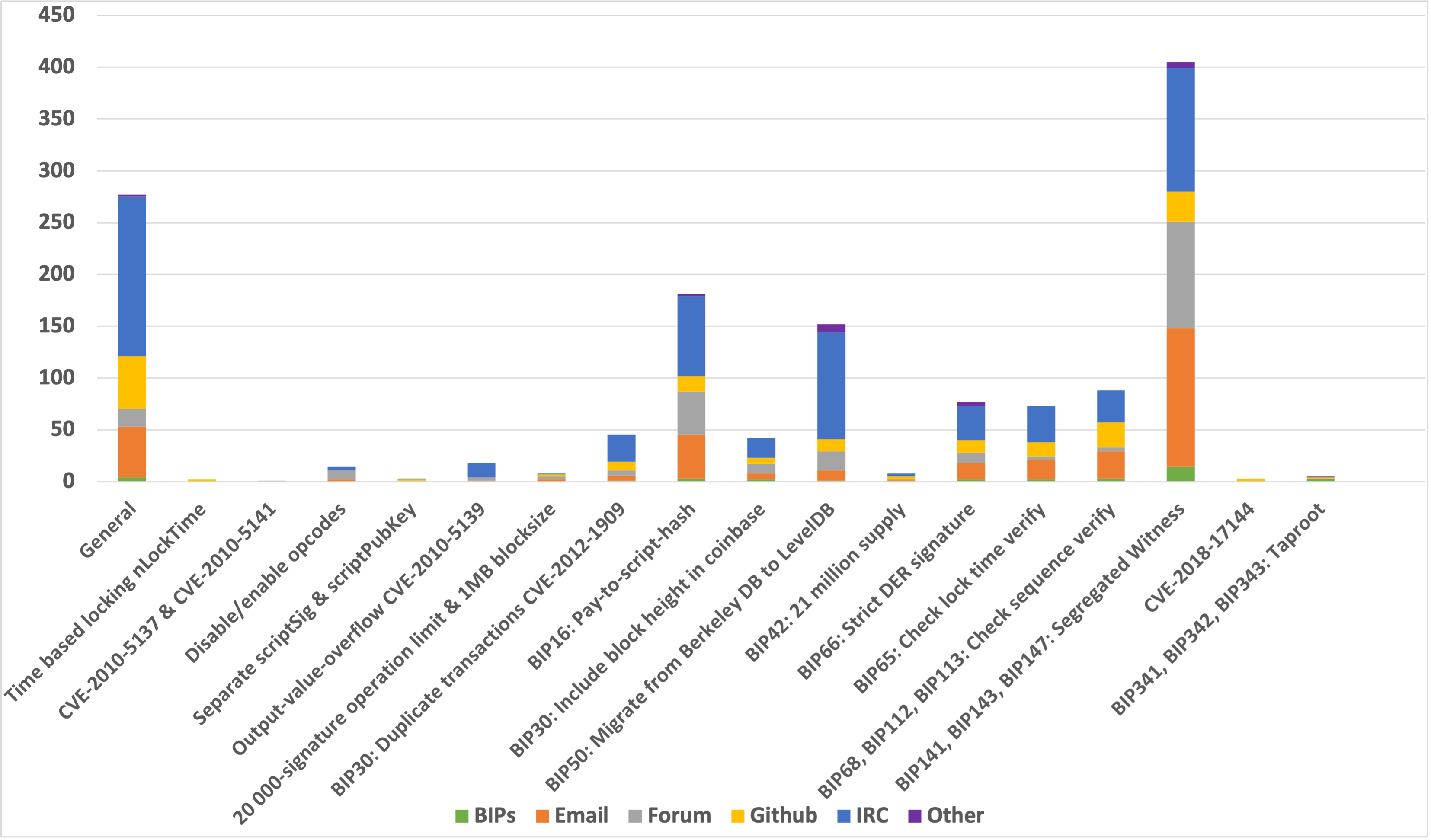}
  \caption{The chart illustrates the distribution of samples tied to consensus change events. The "general" category shows samples tied to overall discussions on consensus change. The remaining bars correspond to changes in Bitcoin's history, suggesting early changes were not extensively discussed under Satoshi's leadership. Post Satoshi's departure (BIP 30 onwards), community engagement increased. The chart also highlights a limited sample rate for the two latest incidents as we reached theoretical saturation. The total sample number across all events exceeds the number of sources presented in the previous chart because some samples are related to multiple events.}
  \label{fig:samplesEvents}
\end{figure*}

\begin{figure*}[t]
  \centering
  \includegraphics[width=0.9\linewidth]{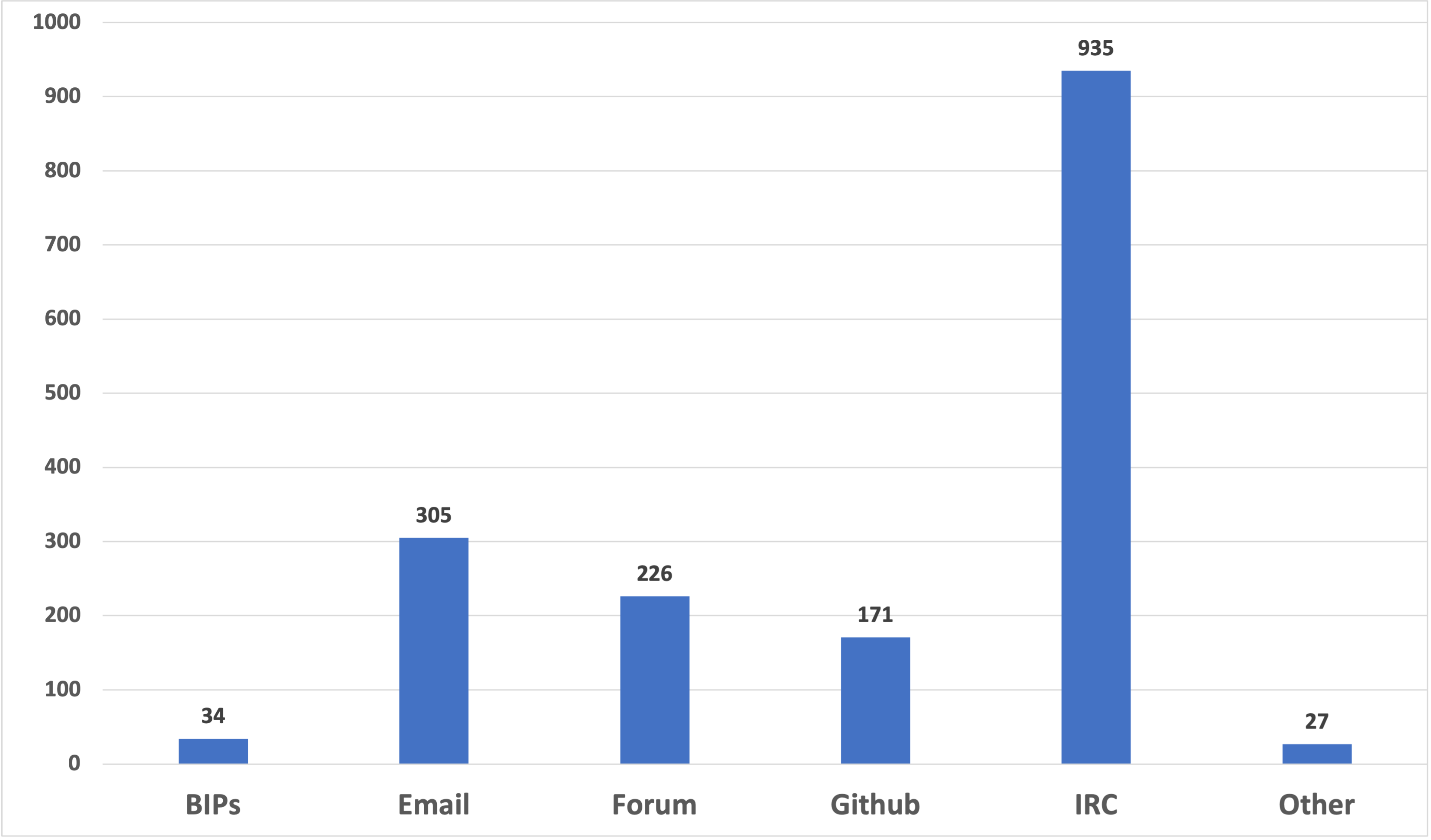}
  \caption{We scraped and filtered samples from different sources and ended up with 1700. The diagram shows the distribution of samples over sources. Samples are counted by 1) A Bitcoin improvement proposal (BIP) specification. 2) An email thread. 3) A forum thread. 4) A Github thread which discusses an issue or a pull request. 5) A whole day of messages from an IRC chat room. 6) Other samples include blog posts, announcements, or videos.}
  \label{fig:samplesSources}
\end{figure*}

\begin{figure*}[t]
    \centering
    \includegraphics[width=1\textwidth]{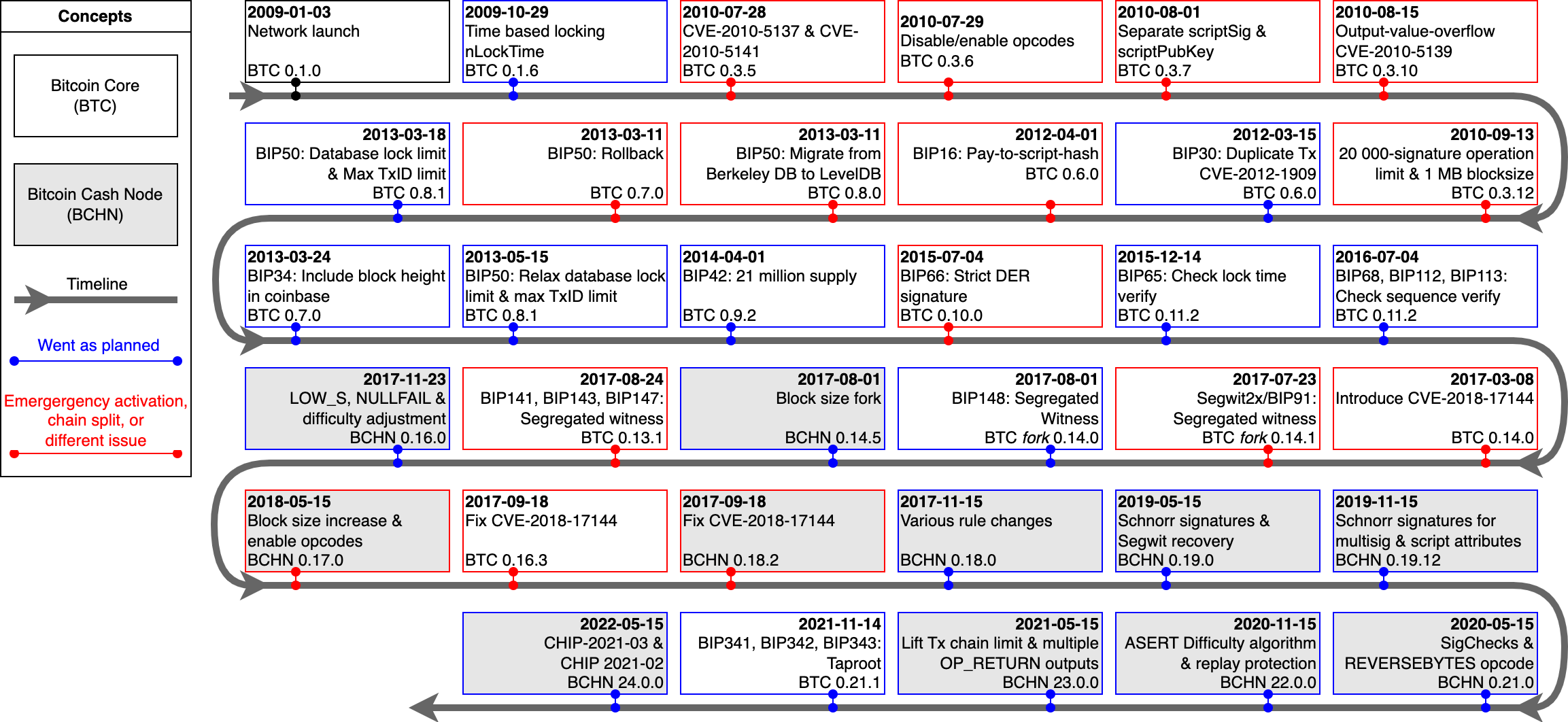}
    \caption{The timeline summarises 34 consensus-changing events until 2023 as first presented in~\cite{self}. It includes 24 changes in BTC and 10 changes in BCH (represented by Bitcoin Cash Node (BCHN)) marked as grey boxes. The dates are based on either the flag day/block for activation, the date the changes activated based on signal thresholds, or the release date. The red boxes indicate issues where the deployment was performed by emergency, caused a chain split, etc. All the events in the timeline represent instances of governing Bitcoin, either by planned changes or handling emergencies.}
    \label{fig:timeline}
\end{figure*}

\clearpage
\section{Demployment signals \& activation timings} \label{signals}

\begin{figure*}
    \centering
    \includegraphics[angle=90, height=0.95\textheight]{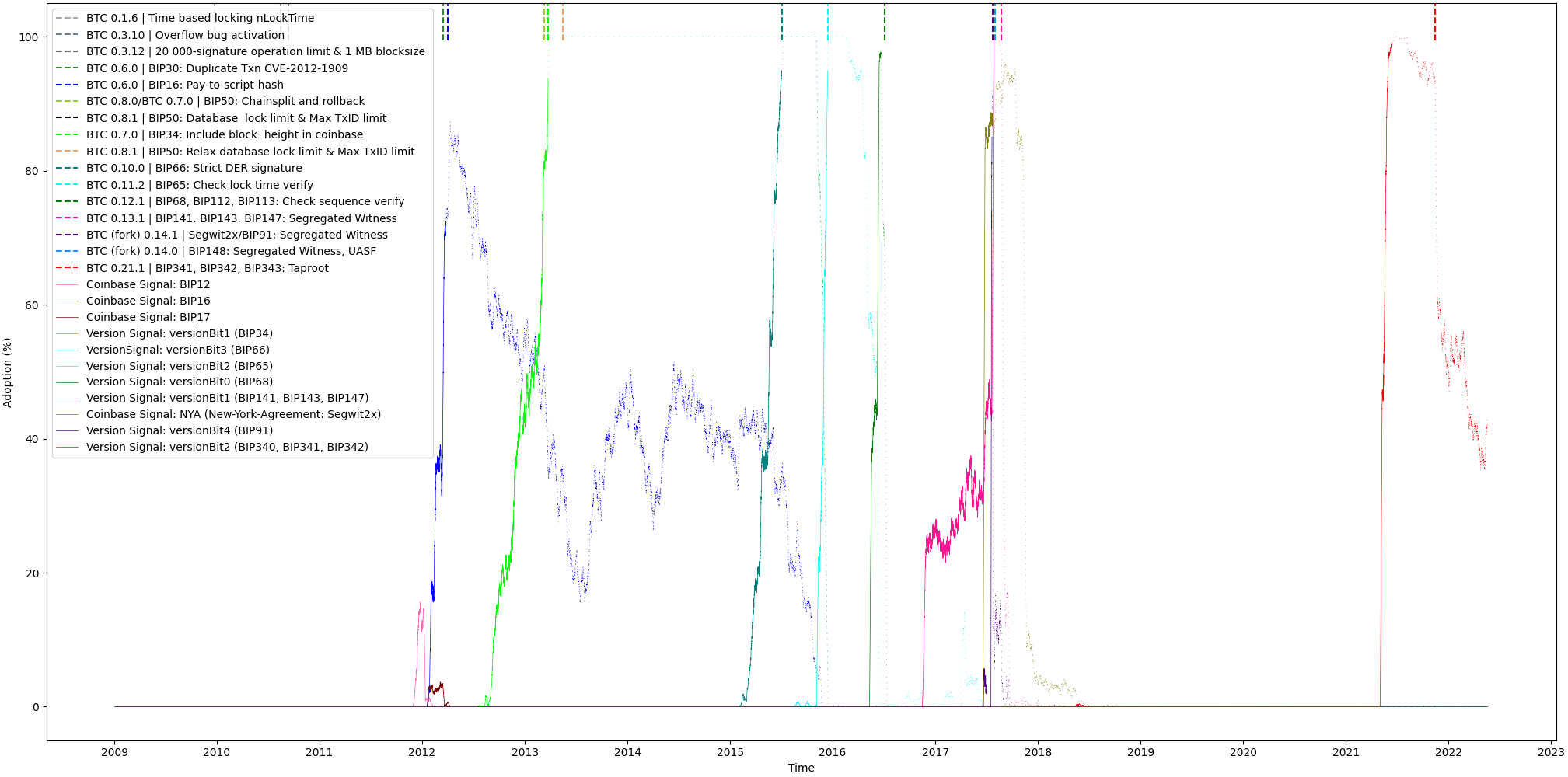}
    \caption{Evolution of BTC including signals and flag day/block activation timings on the Bitcoin chain (BTC)}
    \label{fig:BTCevolution}
\end{figure*}

\begin{figure*}
    \centering
    \includegraphics[angle=90, height=0.95\textheight]{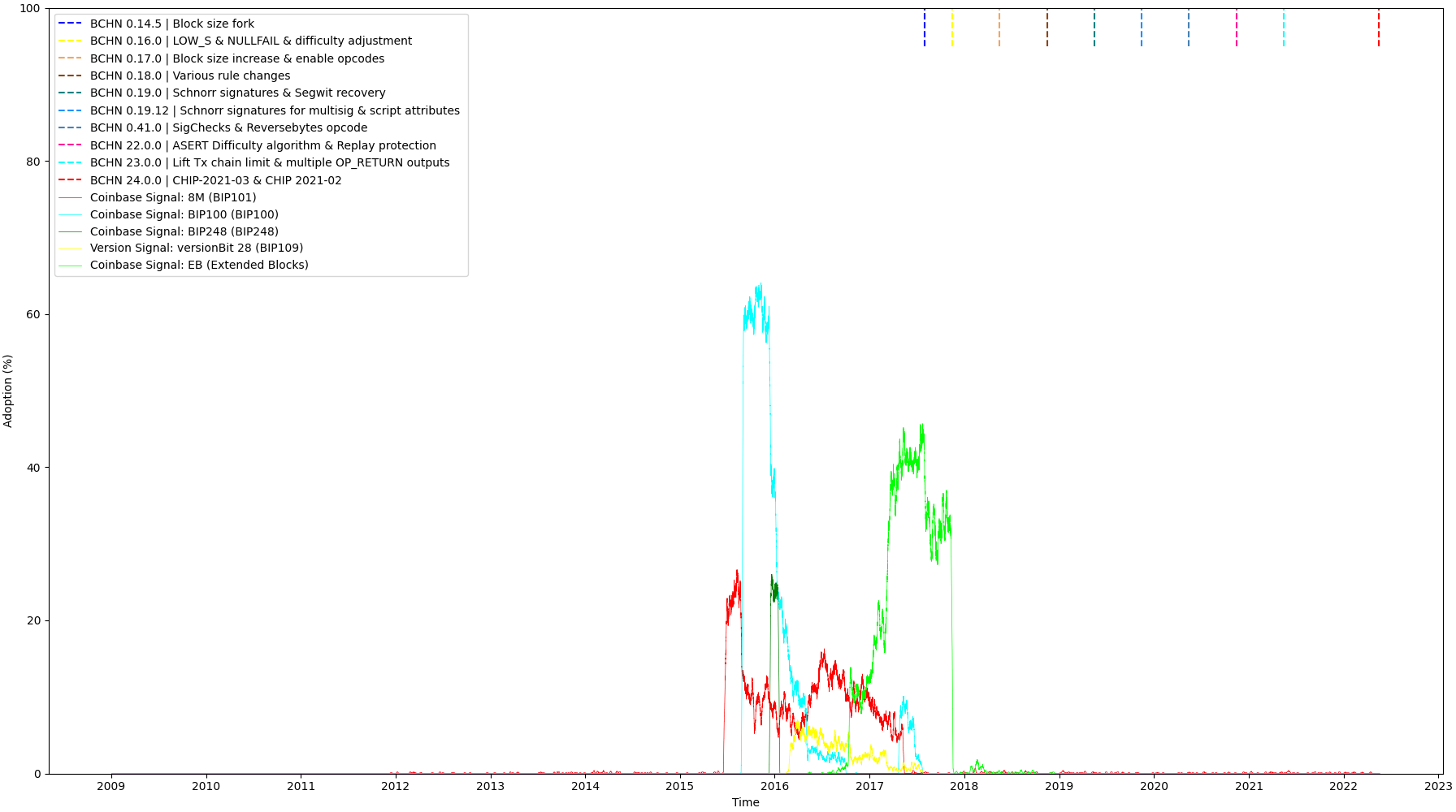}
    \caption{Evolution of BCH, including failed signalling attempts to increase the block size on the Bitcoin chain (BTC) and flag day/block activation timings on the Bitcoin Cash chain (BCH).}
    \label{fig:BCHevolution}
\end{figure*}

We included all the signals used in successful and failed attempts to consensus rules for the Bitcoin blockchain in~\cref{fig:BTCevolution,,fig:BCHevolution}. We kept on including signals from the Bitcoin (BTC) fork after the hard fork, while Bitcoin Cash does not use signals for consensus changes. The on-chain signals and activation timings visualise the evolution and on-chain governance of consensus change events. All the signals originate from the same chain. However, they are shown in two figures to improve readability.

The signals are visualised as graph lines that remain solid until reaching the activation threshold. The dashed lines on top of the graph symbolise the activation of new rules. Thereafter, the graph lines are faded to indicate that consensus nodes no longer consider them.

Note that the block size hard fork is the first consensus rule activation indicated by a dashed line in~\cref{fig:BCHevolution} (BCHN 0.14.5). Therefore, all the activation timings shown in~\cref{fig:BCHevolution} happened on the Bitcoin Cash fork.

%
%
%
\end{document}